\documentclass[twocolumn]{aastex631}

\usepackage{graphicx} 
\usepackage{amsmath}

\shorttitle{Mergers in CEERS Simulated Images}
\shortauthors{Rose et al.}

\graphicspath{{./}}

\begin{document}

\title{Identifying Galaxy Mergers in Simulated CEERS NIRCam Images using Random Forests}

\email{crr9508@rit.edu}

\author[0000-0002-8018-3219]{Caitlin Rose}
\author[0000-0001-9187-3605]{Jeyhan S. Kartaltepe}
\affil{Laboratory for Multiwavelength Astrophysics, School of Physics and Astronomy, Rochester Institute of Technology, 84 Lomb Memorial Drive, Rochester, NY 14623, USA}

\author{Gregory F. Snyder}
\affiliation{Space Telescope Science Institute,
3700 San Martin Dr,
Baltimore, MD 21218, USA}

\author[0000-0002-9495-0079]{Vicente Rodriguez-Gomez}
\affiliation{Department of Physics and Astronomy,
Johns Hopkins University,
Baltimore, MD 21218, USA}
\affiliation{Instituto de Radioastronom\'{i}a y Astrof\'{i}sica,
Universidad Nacional Aut\'{o}noma de M\'{e}xico,
Apdo. Postal 72-3,
58089 Morelia, Mexico}

\author[0000-0003-3466-035X]{L. Y. Aaron\ Yung}
\affiliation{Astrophysics Science Division, NASA Goddard Space Flight Center, 8800 Greenbelt Rd, Greenbelt, MD 20771, USA}

\author[0000-0002-7959-8783]{Pablo Arrabal Haro}
\affiliation{NSF's National Optical-Infrared Astronomy Research Laboratory, 950 N. Cherry Ave., Tucson, AZ 85719, USA}

\author[0000-0002-9921-9218]{Micaela B. Bagley}
\affiliation{Department of Astronomy, The University of Texas at Austin, Austin, TX, USA}

\author[0000-0003-2536-1614]{Antonello Calabr{\`o}} 
\affiliation{Osservatorio Astronomico di Roma, via Frascati 33, Monte Porzio Catone, Italy}

\author[0000-0001-7151-009X]{Nikko J. Cleri}
\affiliation{Department of Physics and Astronomy, Texas A\&M University, College Station, TX, 77843-4242 USA}
\affiliation{George P.\ and Cynthia Woods Mitchell Institute for Fundamental Physics and Astronomy, Texas A\&M University, College Station, TX, 77843-4242 USA}

\author[0000-0003-1371-6019]{M. C. Cooper}
\affiliation{Department of Physics \& Astronomy, University of California, Irvine, 4129 Reines Hall, Irvine, CA 92697, USA}

\author[0000-0001-6820-0015]{Luca Costantin}
\affiliation{Centro de Astrobiolog\'ia (CSIC-INTA), Ctra de Ajalvir km 4, Torrej\'on de Ardoz, 28850, Madrid, Spain}

\author[0000-0002-5009-512X]{Darren Croton}
\affiliation{Centre for Astrophysics \& Supercomputing, Swinburne University of Technology, Hawthorn, VIC 3122, Australia}
\affiliation{ARC Centre of Excellence for All Sky Astrophysics in 3 Dimensions (ASTRO 3D)}

\author[0000-0001-5414-5131]{Mark Dickinson}
\affiliation{NSF's National Optical-Infrared Astronomy Research Laboratory, 950 N. Cherry Ave., Tucson, AZ 85719, USA}

\author[0000-0001-8519-1130]{Steven L. Finkelstein}
\affiliation{Department of Astronomy, The University of Texas at Austin, Austin, TX, USA}

\author[0000-0002-1857-2088]{Boris H\"au{\ss}ler}
\affiliation{European Southern Observatory, Alonso de Cordova 3107, Casilla 19001, Santiago, Chile}

\author[0000-0002-4884-6756]{Benne W. Holwerda}
\affil{Physics \& Astronomy Department, University of Louisville, 40292 KY, Louisville, USA}

\author[0000-0002-6610-2048]{Anton M. Koekemoer}
\affiliation{Space Telescope Science Institute, 3700 San Martin Dr.,
Baltimore, MD 21218, USA}

\author[0000-0002-8816-5146]{Peter Kurczynski}
\affiliation{Observational Cosmology Laboratory (Code 665), NASA Goddard Space Flight Center, Greenbelt, MD 20771, USA}

\author[0000-0003-1581-7825]{Ray A. Lucas}
\affiliation{Space Telescope Science Institute, 3700 San Martin Drive, Baltimore, MD 21218, USA}

\author{Kameswara Bharadwaj Mantha}
\affiliation{Minnesota Institute for Astrophysics, University of Minnesota, 116 church St SE, Minneapolis, MN, 55455, USA.}

\author[0000-0001-7503-8482]{Casey Papovich}
\affiliation{Department of Physics and Astronomy, Texas A\&M University, College Station, TX, 77843-4242 USA}
\affiliation{George P.\ and Cynthia Woods Mitchell Institute for Fundamental Physics and Astronomy, Texas A\&M University, College Station, TX, 77843-4242 USA}

\author[0000-0003-4528-5639]{Pablo G. P\'erez-Gonz\'alez}
\affiliation{Centro de Astrobiolog\'{\i}a (CAB/CSIC-INTA), Ctra. de Ajalvir km 4, Torrej\'on de Ardoz, E-28850, Madrid, Spain}

\author[0000-0003-3382-5941]{Nor Pirzkal}
\affiliation{ESA/AURA Space Telescope Science Institute}

\author[0000-0002-6748-6821]{Rachel S. Somerville}
\affiliation{Center for Computational Astrophysics, Flatiron Institute, 162 5th Avenue, New York, NY, 10010, USA}

\author[0000-0002-4772-7878]{Amber N. Straughn}
\affiliation{Astrophysics Science Division, NASA Goddard Space Flight Center, 8800 Greenbelt Rd, Greenbelt, MD 20771, USA}

\author[0000-0002-8224-4505]{Sandro Tacchella}
\affiliation{Kavli Institute for Cosmology, University of Cambridge, Madingley Road, Cambridge, CB3 0HA, UK}\affiliation{Cavendish Laboratory, University of Cambridge, 19 JJ Thomson Avenue, Cambridge, CB3 0HE, UK}

\begin{abstract}

Identifying merging galaxies is an important -- but difficult -- step in galaxy evolution studies. We present random forest classifications of galaxy mergers from simulated JWST images based on various standard morphological parameters. We describe (a) constructing the simulated images from IllustrisTNG and the Santa Cruz SAM, and modifying them to mimic future CEERS observations as well as nearly noiseless observations, (b) measuring morphological parameters from these images, and (c) constructing and training the random forests using the merger history information for the simulated galaxies available from IllustrisTNG. The random forests correctly classify $\sim60\%$ of non-merging and merging galaxies across $0.5 < z < 4.0$. Rest-frame asymmetry parameters appear more important for lower redshift merger classifications, while rest-frame bulge and clump parameters appear more important for higher redshift classifications. Adjusting the classification probability threshold does not improve the performance of the forests. Finally, the shape and slope of the resulting merger fraction and merger rate derived from the random forest classifications match with theoretical Illustris predictions, but are underestimated by a factor of $\sim 0.5$.

\end{abstract}

\section{Introduction} \label{sec:intro}
Mergers are known to play an important role in the evolution of galaxies over cosmic time. The gravitational interactions between merging galaxies compress gas and create shocks, inducing star formation throughout, and can funnel gas toward their centers, powering nuclear starbursts and fueling active galactic nuclei (AGN) \citep[e.g.,][]{sanders1988a,mihos1996,hopkins2008}. This process can also disrupt the orderly rotation of disk stars, driving the morphological transition of galaxies by turning spiral disk galaxies into ellipticals \citep[e.g.,][]{too1977,cox2006,kor2009,rod2017} as well as inducing disk-instabilites that may cause the build up of the most massive structures at $z > 3$ \citep[e.g.,][]{tacchella2015,zolotov2015, Costantin2021, Costantin2022}. We now know that the rate at which mergers occur evolves strongly out to $z\sim1.5$, as seen by many observational studies as well as cosmological simulations \citep[e.g.,][]{kart2007,bri2007,lotz2011,rod2015,man2018} and that interactions and mergers can have a large impact on the star formation rates and AGN activity of galaxies \citep[e.g.,][]{ell2008,pat2011,lar2016}. 

Studies of the merger rate during cosmic noon ($z\sim 1-3$) have benefited from deep NIR surveys of galaxies with the Wide Field Camera (WFC3) on the {\it Hubble Space Telescope} (HST), though many uncertainties in the observations and tension with expectations from simulations remain. At these higher redshifts, the empirical merger rates of \cite{lotz2011} and \cite{oleary2021} and the theoretical rates of \cite{hop2010} and \cite{rod2015} continue to increase (albeit at different rates). On the other hand, \cite{man2016} find that their empirical merger rate flattens, while \cite{man2018} find that their empirical rate either turns over and begins decreasing, or remains increasing, depending on the criteria used for their merger selection. 

\cite{dun2019} compare their observed merger rates from the Cosmic Assembly Near-infrared Deep Extragalactic Legacy Survey \citep[CANDELS; ][]{koe2011,gro2011} to previous studies. They find that for galaxies with $\log_{10} (M_{\star} / M_{\odot}) > 10.3$, their increasing merger rate agrees with that measured in the Illustris simulation by \cite{rod2015} out to $z\sim6$, as well as with \cite{mun2017} out to $z\sim2$. However, beyond $z\sim2$, \cite{dun2019} disagrees with the rate from \cite{man2018} which begins decreasing. For galaxies with $9.7 < \log_{10} (M_{\star} / M_{\odot}) < 10.3$, the increasing rate of \cite{dun2019} agrees with that of \cite{ven2017} out to $z\sim3$ (within their error bars), yet disagrees with the shallower rate from Illustris, indicating that the Illustris merger rates are in tension with observations. Furthermore, \cite{dun2019} find that their comoving merger rates suffer from significant uncertainties at $z > 4$. Additionally, \cite{fer2020} identify mergers using deep learning and find that the rates broadly agree with visual classifications out to $z\sim3$.

There are several sources of uncertainty that currently plague our understanding of mergers at these high redshifts. First, modern surveys with WFC3 are only sensitive to the most massive galaxies and major mergers \citep[mass ratio $<$4:1; e.g.,][]{Ellison2013,man2018} and the numbers detected drop-off sharply beyond $z>3$. At these redshifts, it is expected that low mass galaxies and minor mergers (mass ratios between 4:1 and 10:1) may play an increasing role \citep[e.g.,][]{Kaviraj2014}. Furthermore, the conversion of an observed merger fraction into a merger rate requires the assumption of a merger time scale and how it might evolve with redshift. This timescale itself is highly uncertain and relies on information from simulations \citep{sny2017, dun2019}. Finally, identifying mergers from images in the first place poses many of its own challenges.

Typically, there have been two methods employed to identify merging systems: 1) identifying close pairs of galaxies that are likely to merge at some point in the future  and 2) identifying advanced mergers through morphological disturbances, such as double nuclei, tidal tails, and other asymmetries. The close pair method finds galaxy pairs that are close on the sky and in redshift, using either spectroscopic redshifts \citep[e.g.,][]{lin2004,tas2014,ven2017,shah2020} or a sophisticated analysis of photometric redshifts \citep[e.g.,][]{kart2007,man2018,dun2019}. Identifying galaxy mergers through morphological features can be done via visual classifications \citep[e.g.,][]{lin2008, kart2015}, and is typically robust since the human eye is skilled at picking out patterns and faint features in noisy images. However, visual classifications can be both subjective and time-consuming, especially for large surveys. Quantitative parameters such as $CAS$, $G$, $M_{20}$, and the $MID$ statistics \citep[e.g.,][]{con2003, lotz2004, free2013} were developed as a less subjective alternative to visual classifications.

In populations of low redshift galaxies, these quantitative morphology parameters have been shown to effectively locate mergers, since they can be correlated with features like asymmetries, multiple cores, starbursts, and tidal tails that are caused by merging \citep[e.g.,][]{con2003,lotz2004,lotz2008,free2013,wen2014,snylotz2015,paw2016}. However, for high redshift galaxies, these parameters become less effective at classifying mergers \citep[e.g.,][]{lotz2004,con2008,kartaltepe2010,thom2015,snylotz2015}. The main reason for this is simply that at high redshifts, mergers are more difficult to see -- cosmological surface brightness dimming leads to the loss of low surface brightness features, poor angular resolution leads to the blurring of small scale structure, and rest-frame optical emission is shifted to the near- and mid-infrared, which means optical, and short-wavelength instruments like ACS and WFC3 on HST are unable to observe the structure of the general stellar population. Furthermore, high redshift galaxies can have intrinsically more clumpy and irregular morphologies than those at low redshifts, which could masquerade as merger signatures \citep[e.g.,][]{Dekel2009,Kartaltepe2012,kart2015}.

\begin{figure}
\includegraphics[scale=0.28]{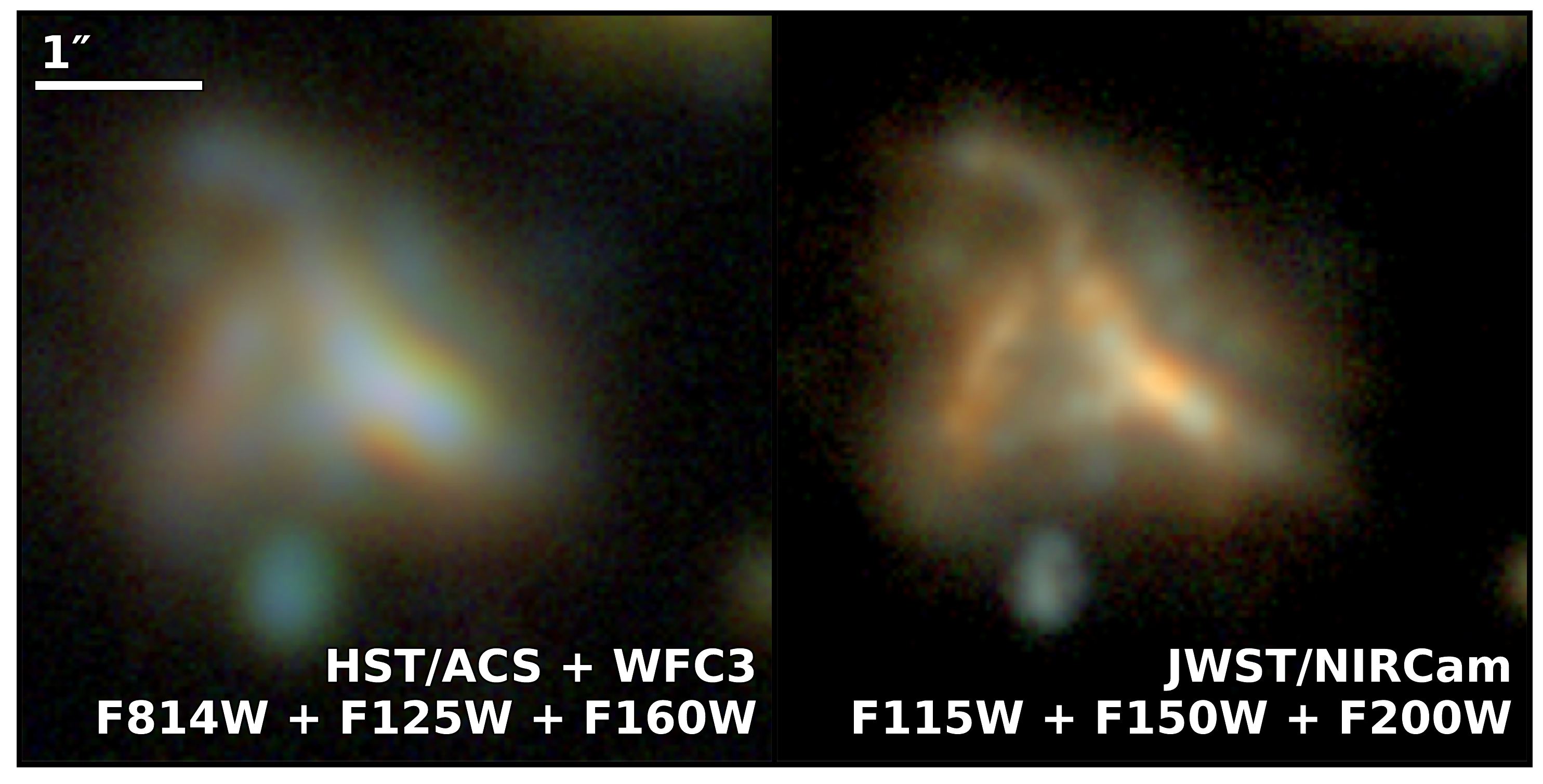}
\caption{RGB images of an IllustrisTNG simulated $z=3$ galaxy (log(M$_{\star} / M_{\odot}$) = 10.47) as seen with HST/ACS+WFC3 (\textit{\textbf{left}}) and with JWST/NIRCam (\textit{\textbf{right}}). See \S \ref{sec:simdata} for details about the IllustrisTNG images used in this work. \label{fig:hst_v_jwst}}
\end{figure}

Therefore, high quality, deep,  high-resolution near-infrared imaging is needed to detect merger features at high redshifts and subsequently curate an accurate sample set of high redshift galaxy mergers. The state-of-the-art and recently launched {\it James Webb Space Telescope} (JWST) will provide the highest quality images of distant galaxies to date. One of the first deep extragalactic surveys that JWST will undertake is the Cosmic Evolution Early Release Science (CEERS) Survey (PI: S. Finkelstein). CEERS utilizes JWST imaging and spectroscopy in parallel over $100$ arcmin$^2$ in the Extended Groth Strip HST legacy field for $0.5 < z < 13$ galaxies. CEERS uses JWST’s near-infrared camera (NIRCam) to probe structural features at higher redshifts than has been possible with HST \citep{fink2017}. Figure \ref{fig:hst_v_jwst} compares the same simulated $z=3$ galaxy as seen with HST/ACS + WFC3 and with JWST/NIRCam using the CEERS observing strategy, at similar wavelengths. The improved resolution of the JWST CEERS images will result in much more accurate morphological measurements than the HST images.

In addition to using deeper, high resolution NIR imaging from JWST, new analysis techniques can be employed to better identify the morphological signatures of mergers. In particular, machine learning methods show great promise for identifying high redshift mergers. At low redshift, a number of studies have already used machine learning for morphological classifications \citep[e.g.,][Guzman-Ortega et al., in preparation]{hue2015,peth2016,sreejith2018,bot2019,nev2019,pearson2019, cheng2020}. For merger classifications in particular, \cite{nev2019} note that their machine learning method outperforms the classical automated identification methods (e.g., \citealt{con2003, lotz2004}) in the nearby universe. Recently, studies have begun to explore using machine learning to identify high redshift mergers \citep{sny2019,cip2020, fer2020,Ferreira2022,sharma2021arXiv}. In particular, \cite{sny2019} use random forests to identify $0.5 < z < 4$ mergers from Illustris-1 and find that their classifier achieves a true positive rate of up to $\sim70\%$ for mergers in simulated HST images. They also note that their classifier generally outperforms the more simple classifiers \`{a} la \cite{con2003} and \cite{lotz2004} across their redshift range.

In this paper, we explore the use of random forests in identifying galaxy mergers from simulated CEERS
images from IllustrisTNG in preparation for new JWST observations. In \S \ref{sec:data}, we describe the simulated JWST images used in this work as well as how those images were modified to create one set that matches CEERS NIRCam observations and other effectively noiseless sets for comparison purposes. In \S \ref{sec:params}, we describe the large set of quantitative morphology parameters used as input to the forests and how they were calculated from the data. In \S \ref{sec:analysis} and \S \ref{sec:discuss}, we present our random forest analyses and discuss the results. We summarize and conclude in \S \ref{sec:con}.

\section{Data} \label{sec:data}
\subsection{CEERS NIRCam Imaging}

CEERS is an Early Release Science program (PI S. Finkelstein) to observe the EGS (Extended Groth Strip; \citealt{davis2007}) extragalactic deep field (one of the five CANDELS fields; \citealt{koe2011,gro2011}) early in Cycle 1. The NIRCam imaging of CEERS will cover 10 pointings over 100 sq.\ arcmin with the F115W, F150W, F200W, F277W, F356W, and F444W filters down to a 5$\sigma$ depth ranging from 28.4–29.2. CEERS NIRCam imaging will enable spatially resolved rest-optical/near-IR measurements for $1 < z < 7$ galaxies, with a resolution of $< 1$ kpc at 3.6 $\mu$m. Table \ref{tab:ETCsetup} presents the CEERS observing strategy and integrated observing time.

CEERS is expected to reveal objects with irregular and perturbed morphologies in great detail, which will allow astronomers to track the structural evolution of galaxies from $z \sim 7$ to today. Since CEERS is targeting the EGS HST field, CEERS data will also be accompanied by a rich set of multi-wavelength data from HST ACS \citep{davis2007}, HST WFC3 \citep{koe2011,gro2011,momcheva2016}, {\it Spitzer} \citep{ashby2013}, {\it Herschel} \citep{lutz2011,oliver2012}, {\it Chandra} \citep{laird2009,nandra2015}, and a number of ground-based imaging and spectroscopic surveys \citep[e.g.,][]{coil2004,cooper2012b,newman2013,kriek2015,masters2019} with high-quality photometric reshifts and stellar masses \citep{stefanon2017}. JWST launched in December 2021, with the first CEERS images observed in June 2022 and released in July 2022, including 4 out of the 10 NIRCam pointings that make up the complete CEERS NIRCam mosaic in the EGS field. The remaining 6 pointings will be observed in December 2022.

\subsection{The Simulated Images} \label{sec:simdata}

We construct mock images of the EGS field from IllustrisTNG100-1, one of three cosmological simulations of galaxy formation and evolution included in the IllustrisTNG suite \citep{TNG1Springel2018,TNG2Naiman2018,TNG3Nelson2018,TNG4Pillepich2018,TNG5Marinacci2018}. Compared to Illustris-1, IllustrisTNG uses several updated prescriptions for physical processes (such as magnetic fields, black hole feedback, and galactic winds) as detailed by \cite{Weinberger2017} and \cite{pill2018}. TNG100-1 has been shown to produce galaxy morphologies in good agreement with observations \citep[e.g.,][]{rod2019,Tacchella2019}.

To construct simulated images, we first adopt a simulated lightcone with footprint overlapping the observed EGS field, constructed using dark matter halos from an n-body simulation and the Santa Cruz semi-analytic model (SAM) for galaxy formation \citep{som2021, yung2022}. The Santa Cruz SAM tracks a wide variety of baryonic processes using prescriptions derived analytically, inferred from observations, or extracted from numerical simulations, and provides physically-backed predictions for galaxies across wide ranges of redshift and mass. This model has been shown to be able to reproduce the observed evolution in distribution functions of rest-frame UV luminosity, stellar mass, and SFR from $z \sim 0$ to the highest redshift where observational constraints are available \citep{som2015,yung2019_1,Yung2019}.

\begin{figure*}
\plotone{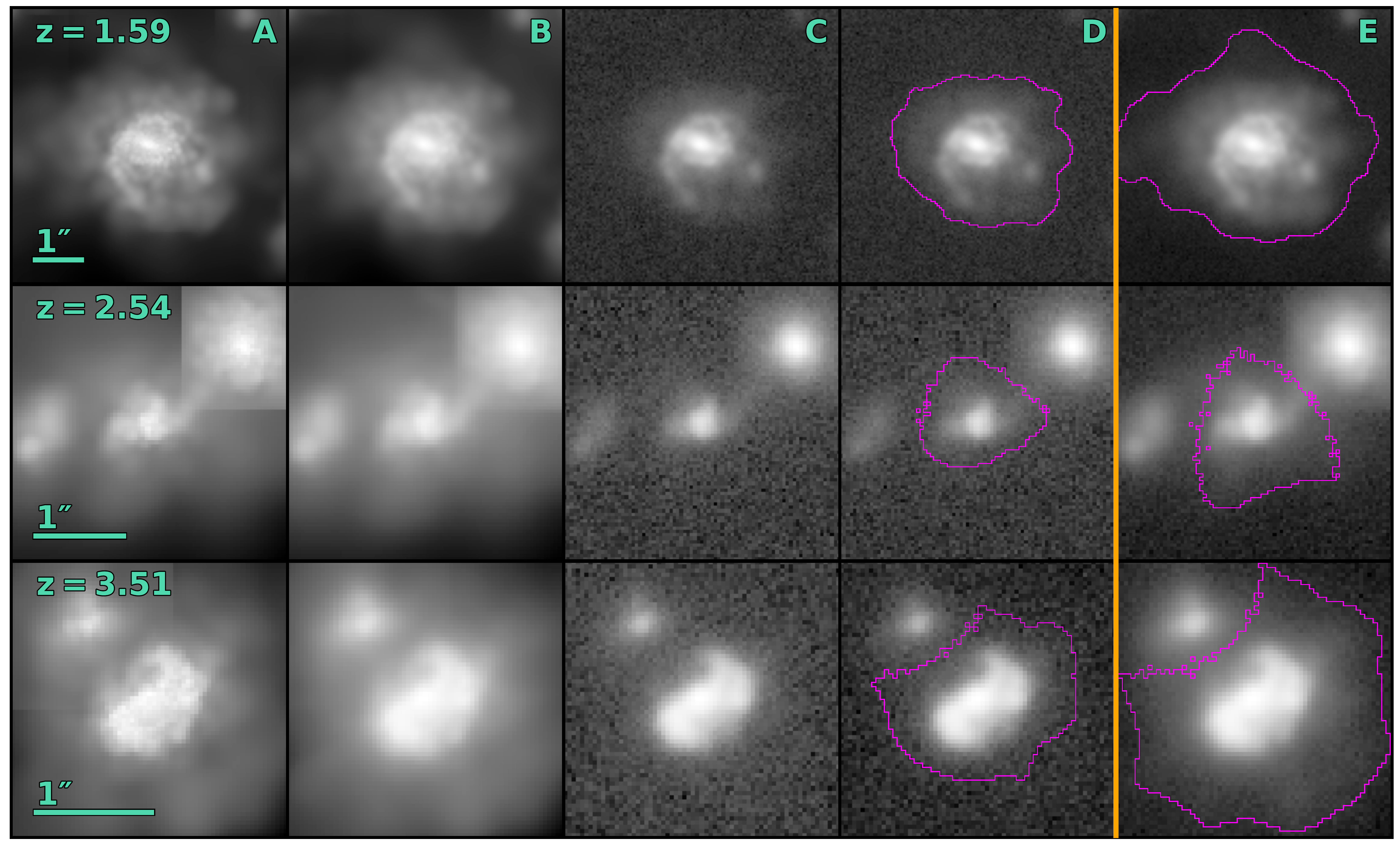}
\caption{Examples of galaxy mergers at different redshifts from the original F277W image (\textbf{column A}) before undergoing the following modifications: convolution with the PSF (\textbf{column B});
inclusion of Poisson noise and CEERS background noise (\textbf{column C}); and then background subtraction (\textbf{column D}). The nearly noiseless versions are shown in \textbf{column E}. The magenta contours in columns D and E are the segmentation map outlines from \texttt{Source Extractor}.
\label{fig:stamps}}
\end{figure*}

Then, for each galaxy in the SAM, we identify IllustrisTNG subhalos from the TNG100-1 simulation by searching in the space of halo mass versus star formation rate. We randomly choose matching subhalos from those within a factor of two in these dimensions from the SAM galaxy catalogue. We then create simulated images of each subhalo using the public visualization API \citep{Nelson2019} in each of the JWST NIRCam F115W, F150W, F200W, F277W, F356W, and F444W filters, and add them together to form the wide-area mock images. These images cover an area of $\sim$100 square arcmin, to mimic the size of the EGS field that CEERS will cover, and contain over 100,000 galaxies. The pixel scale of the images is 0.03 arcsec/pix.

These images are accompanied by catalogs with intrinsic information such as redshift, star formation rate, and stellar mass. Additionally, the merger history catalogs for IllustrisTNG galaxies \citep{rod2015,nel2019} used in this work give the IllustrisTNG snapshot numbers for each galaxy's most recent past merger and next future merger (both major and minor), as well as the total number of past mergers experienced in the galaxy's history for different timescales. The merger history information available for these galaxies makes IllustrisTNG an ideal dataset for training and testing machine learning algorithms to identify galaxy mergers at different stages.

\subsection{Modifying the Simulated Images}
\begin{table*}[t]
\centering
\begin{tabular}{lcccccc}
\hline
\hline
 & Subarray & Readout pattern & Groups per & Integrations & Exposure per & Exposure time \\
 & & & integration & per exposure & specification & [sec] \\
\hline
CEERS & FULL & DEEP8 & 5 & 3 & 1 & 2855.98 \\
Effectively Noiseless & FULL & DEEP8 & 80 & 60 & 1 & 1023632.92 \\
\hline
\end{tabular}
\caption{JWST ETC detector setup and resulting exposure times for each set of images. The CEERS setup here is described in the original CEERS proposal \citep{fink2017}. This setup has since changed to MEDIUM8 with 9 groups times 3 exposures for the actual CEERS observations. However, the final exposure times are similar and should not affect our results.}
\label{tab:ETCsetup}
\end{table*}

We start with the pristine simulated images, and convolve them with the PSF of each filter, which are the model PSFs from WebbPSF \citep{perrin2014}.
We then add Poisson noise (due to the galaxies in the image) following the formulation of \cite{pont2016}, where each image is convolved with a kernel to sum fluxes in neighboring pixels. We then add background noise to represent the actual CEERS exposure times (see Table \ref{tab:ETCsetup}), which was estimated using the exposure time calculator (ETC) system developed for JWST \citep{pont2016}. Last, the Python package \texttt{photutils} is used to estimate the average background in each image, which is then subtracted from the images resulting in a final set of CEERS-like, background subtracted images for each of the six NIRCam filters. Figure \ref{fig:stamps} shows  examples of simulated mergers at each step of the image modification process.

\begin{table*}[t]
\centering
\begin{tabular}{lcccccc}
\hline
\hline
& \texttt{DETECT\_MINAREA} & \texttt{DETECT\_THRESH} & \texttt{ANALYSIS\_THRESH} & \texttt{DEBLEND\_NTHRESH} & \texttt{DEBLEND\_MINCONT} \\
& cold / hot & cold / hot & cold / hot & cold / hot & cold / hot \\
\hline
CEERS & 10 / 15 & 4.5 / 1.7 & 5.0 / 0.8 & 32 / 64 & 0.01 / 0.001 \\
Effectively Noiseless & 10 / 15 & 4.5 / 1.7 & 5.0 / 0.8 & 32 / 64 & 0.001 / 0.0001\\
\hline
\end{tabular}
\caption{\texttt{Source Extractor} parameters for each set of images.}
\label{tab:se}
\end{table*}

We similarly create a set of effectively noiseless images, using an extremely long exposure time of 11 days in the ETC (see Table \ref{tab:ETCsetup}), so that we can compare measurements from our CEERS-like images to those from the noiseless ones. This step is necessary because a purely noiseless set of images causes the \texttt{Galapagos-2} code (see \S \ref{sec:params}) to output unusually large errors, and therefore true noiseless images cannot be used for our analysis. Examples of nearly noiseless galaxies are shown in column E of Figure \ref{fig:stamps}. Finally, the headers of each image are amended to include the keywords required by \texttt{Galapagos-2}. 

All of our final CEERS-like and noiseless simulated NIRCam images are available to the public via \href{https://ceers.github.io/releases.html}{https://ceers.github.io/releases.html}.

\section{Quantitative Morphology Parameters} \label{sec:params}

This work makes use of several different morphology parameters. Parametric measurements assume that the galaxy's light distribution follows a specific mathematical profile, such as the S\'ersic profile, specified by the S\'ersic index $n$ \citep{ser1963} and other parameters. Nonparametric measurements do not assume an underlying mathematical form, but rather are statistical measures of the light distribution in a galaxy (e.g., the $CAS$ parameters; \citealt{ber2000,con2000,con2003}).

S\'ersic parameters were calculated using the IDL program \texttt{Galapagos-2} from the \texttt{MegaMorph} Project\footnote{\href{https://www.nottingham.ac.uk/astronomy/megamorph/}{https://www.nottingham.ac.uk/astronomy/megamorph/}\label{fnote:mm}}. \texttt{MegaMorph} is a project designed to improve astronomers' ability to measure the structure of galaxies via parametric methods while making full use of modern multiwavelength imaging surveys \citep{bam2011,hau2013,vika2013}. Using multiwavelength information allows one to constrain fit parameters that vary smoothly as a function of wavelength, which produces more physically consistent models. Under the \texttt{MegaMorph} Project, \texttt{Galapagos}\footnote{\href{https://borishaeussler.github.io/galapagos\_v1/home.html}{https://borishaeussler.github.io/galapagos\_v1/home.html}} \citep{bar2012} was modified to create \texttt{Galapagos-2}, and \texttt{Galfit}\footnote{\href{https://users.obs.carnegiescience.edu/peng/work/galfit/galfit.html}{https://users.obs.carnegiescience.edu/peng/work/galfit/galfit.html}} \citep{peng2002,peng2010} was modified to create \texttt{GalfitM}.

\texttt{GalfitM} is a least-squares fitting algorithm that finds the optimum solution to the S\'ersic fit for a galaxy \citep{peng2002,peng2010,peng2012}. \texttt{Galapagos} (Galaxy Analysis over Large Areas: Parameter Assessment by \texttt{Galfit}-ting Objects from \texttt{SExtractor}) is essentially an \texttt{IDL} wrapper routine that allows \texttt{GalfitM} to be used for large survey images. In addition to the final output catalog with the S\'ersic fit parameters, \texttt{Galapagos-2} also outputs the original stamp, the \texttt{GalfitM} model, and the residual image for each galaxy detected in the survey images. \texttt{Galapagos-2}/\texttt{GalfitM} can also do bulge-disk decomposition and output separate S\'ersic parameters for both galaxy components.

\texttt{Galapagos-2} uses the \texttt{Source Extractor} program for object detection \citep{ber1996}. \texttt{Galapagos-2} first runs \texttt{Source Extractor} in ``cold" mode to deblend nearby galaxies, then in ``hot" mode to detect faint galaxies. The ``cold" and ``hot" catalogues are then combined. Table \ref{tab:se} lists our ``cold" and ``hot" mode parameters for both sets of images.

The Python package \texttt{statmorph}\footnote{\href{https://statmorph.readthedocs.io/en/latest/}{https://statmorph.readthedocs.io/en/latest/}} was used for calculating nonparametric morphology measurements as well as single S\'ersic fits \citep{rod2019}. The inputs to \texttt{statmorph} are the science image, the segmentation map created by \texttt{Source Extractor}, the PSF, and the gain. While \texttt{statmorph} can handle both single galaxy images or survey images, it cannot handle multiple images at once to make full use of multiwavelength data. The morphology measurements from \texttt{statmorph} include: concentration ($C$), asymmetry ($A$), and clumpiness/smoothness ($S$) \citep{ber2000,con2000,con2003}; the Gini coefficient ($G$) and the moment of light ($M_{20}$) \citep{abra2003,lotz2004}; multimode ($M$), intensity ($I$), and deviation ($D$) \citep{free2013}; and  outer asymmetry ($A_O$) and shape asymmetry ($A_S$) \citep{wen2014,paw2016}, as well as various parameters such as radii (e.g., $r_{20}$, $r_{50}$, $r_{80}$, $r_{\textrm{half}}$, $r_{\textrm{petro}}$) and signal-to-noise per pixel.

We also make use of the residual images output by \texttt{GalfitM}. Residual images have been shown to have the potential to highlight asymmetric or unusual structures not captured by the S\'ersic model \citep[e.g.,][]{man2019}. We run \texttt{statmorph} on all residual images to obtain residual morphology measurements. Since \texttt{statmorph} requires that images have positive flux, which was not true for all residual images, we add an offset of $+1$ to every pixel of all residual images. This addition does affect \texttt{statmorph}'s measurements, but will be consistent across all residual images, both mock CEERS and nearly noiseless.

\section{Merger Identification} \label{sec:analysis}

As described in \S \ref{sec:intro}, machine learning techniques are potentially more advantageous for high redshift merger identification compared to classical techniques, due to their ability to exploit complex multidimensional information. Here, we choose to explore the random forest technique \citep{ho1995,brei2001}, which we describe in \S \ref{sec:RFexp}. Prior to implementing random forests, we first prepare the morphology catalog that will be the features input to the forests (\S \ref{sec:catalog}) and define the merger and non-merger classes that will be the labels input to the forests (\S \ref{sec:mergerdef}). We also show an example of the performance of a classical method using our mock CEERS dataset, which further illustrates the need for machine learning merger identification (\S \ref{sec:gm20}).

\subsection{Catalog Creation} \label{sec:catalog}
The IllustrisTNG merger history catalog \citep{rod2015,nel2019} contains IllustrisTNG snapshot numbers for each galaxy, as well as snapshot numbers for the most recent and next merger events, for both major and minor mergers. We convert the snapshot numbers to ages of the universe in Gyr, then calculate the difference in time between the galaxy of interest and the last and next merger events. The resulting merger history catalog therefore contains the time since the last merger (both major and minor) and the time until the next (both major and minor).

After running \texttt{Galapagos-2} and \texttt{statmorph}, we combine these catalogs with the merger history catalog using \texttt{match\_coordinates\_sky()} from the \texttt{astropy} Python package. The final master catalog contains 109,395 galaxies, although only $\sim$70,000 are detected by \texttt{Source Extractor} in the simulated CEERS images and subsequently have morphology measurements. The final catalog therefore contains columns for intrinsic information such as redshift and merger timescales, as well as the \texttt{Galapagos-2} and \texttt{statmorph} measurements.

\begin{figure}
\includegraphics[scale=0.26]{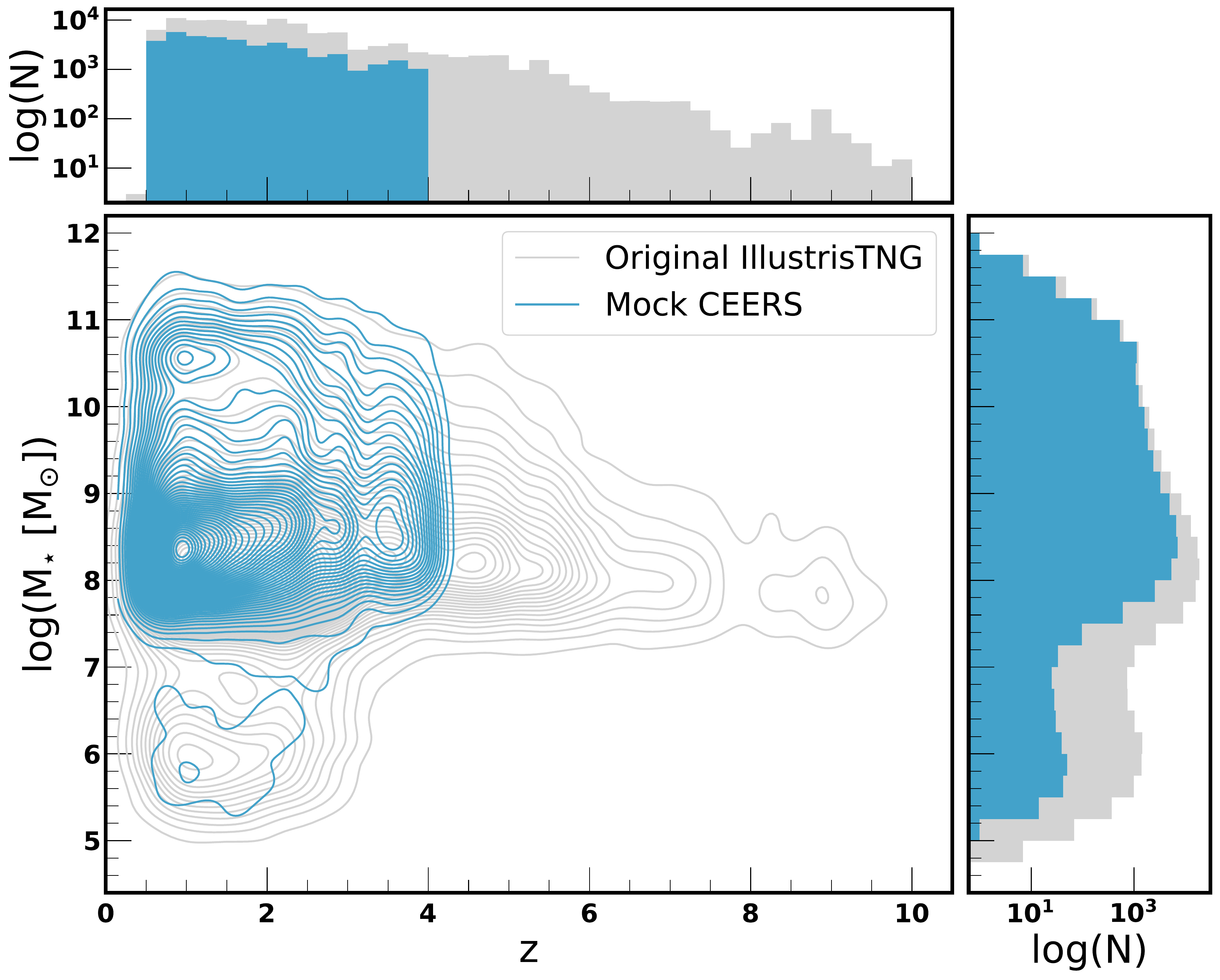}
\caption{Stellar mass versus redshift for objects from the full original IllustrisTNG sample (\textit{grey}), and from the mock CEERS sample with $z<4$, the focus of this paper (\textit{blue}). Above and to the right are the distributions of redshift and stellar mass, respectively.
\label{fig:mvz}}
\end{figure}

\begin{figure*}
\plotone{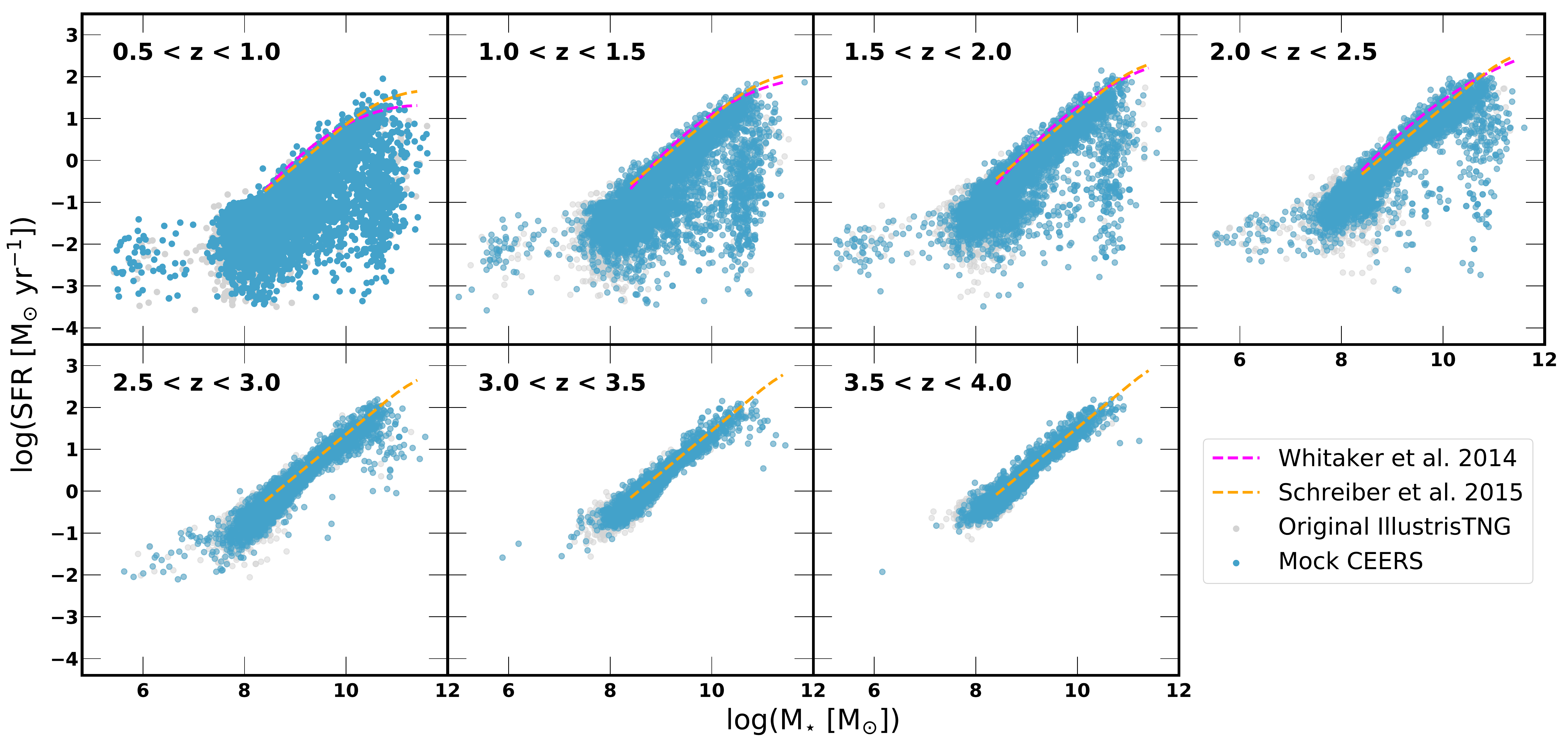}
\caption{Star formation rate versus stellar mass for objects from the full original IllustrisTNG sample (\textit{grey}), and from the $z < 4$ mock CEERS sample (\textit{blue}), split by redshift bin. The main sequence fits from \citet[][\textit{purple}]{whit2014} and \citet[][\textit{orange}]{sch2015} are shown for comparison.
\label{fig:sfrm}}
\end{figure*}

For our analysis, we restrict our mock CEERS dataset to have \texttt{statmorph} signal-to-noise per pixel (S/N$_{F115W}$) $> 3$, as well as \texttt{Flag}$_{\texttt{Galfit}}$ $=2.0$ which indicates successful \texttt{Galfit} measurements. We also restrict our dataset to only galaxies with $0.5\leq z \leq4.0$ to match the redshift range chosen in \cite{sny2019}. Figures \ref{fig:mvz} and \ref{fig:sfrm} illustrate some basic properties of galaxies in the full original IllustrisTNG dataset compared to the galaxies detected in our mock CEERS-like images. Figure \ref{fig:mvz} shows that the full redshift range extends from $z = 0.5$ up to 10, with only a few galaxies existing in the highest redshift bins. It also shows that the stellar masses in our $z < 4$ mock CEERS span a range from $\log(M_{\star} / M_{\odot}) \sim 5 - 12$, but there are far fewer low mass objects in the mock CEERS set than in the original IllustrisTNG set. All of the lowest mass galaxies (both original IllustrisTNG and mock CEERS) are in the lower redshift bins. Figure \ref{fig:sfrm} shows star formation rate as a function of stellar mass, divided into the redshift bins used in this work. The IllustrisTNG data (and mock CEERS data) follows the main sequence fits of \cite{whit2014} and \cite{sch2015}, with a notable lack of starbursts lying above the main sequence.

\begin{table*}
\centering
\begin{tabular}{lccccc}
\hline
\hline
 Redshift Bin & Dataset & Number of Objects & Number of Mergers & Training Set Number & Test Set Number \\
  & & & (major $+$ minor) & \\
\hline
0.5 - 1.0 & CEERS & 9450 & 276 & 6332 & 3119 \\
& EN Set 1 & 9553 & 264 & 6400 & 3153 \\
& EN Set 2 & 7588 & 203 & 5083 & 2505 \\
\hline
1.0 - 1.5 & CEERS & 9227 & 618 & 6182 & 3045 \\
& EN Set 1 & 10937 & 679 & 7327 & 3610 \\
& EN Set 2 & 7748 & 498 & 5191 & 2557 \\
\hline
1.5 - 2.0 & CEERS & 7009 & 827 & 4696 & 2313 \\
& EN Set 1 & 18808 & 3007 & 12601 & 6207 \\
& EN Set 2 & 5939 & 671 & 3979 & 1960 \\
\hline
2.0 - 2.5 & CEERS & 6144 & 1242 & 4116 & 2028 \\
& EN Set 1 & 9395 & 1922 & 6294 & 3101 \\
& EN Set 2 & 5349 & 1086 & 3583 & 1766 \\
\hline
2.5 - 3.0 & CEERS & 3809 & 1174 & 2552 & 1257 \\
& EN Set 1 & 5726 & 1785 & 3836 & 1890 \\
& EN Set 2 & 3317 & 1008 & 2222 & 1095 \\
\hline
3.0 - 3.5 & CEERS & 2199 & 729 & 1473 & 726 \\
& EN Set 1 & 3087 & 1128 & 2068 & 1019 \\
& EN Set 2 & 1920 & 628 & 1286 & 634 \\
\hline
3.5 - 4.0 & CEERS & 2553 & 923 & 1710 & 843 \\
& EN Set 1 & 3470 & 1401 & 2324 & 1146 \\
& EN Set 2 & 2262 & 830 & 1515 & 747 \\
\hline
\end{tabular}
\caption{Dataset sizes for each redshift bin. Major mergers are defined to have a mass ratio less than 4:1 and minor mergers are defined to have a mass ratio between 4:1 and 10:1. The merger timescale is $\pm 0.25$ Gyr.}
\label{tab:data}
\end{table*}

For comparison with the effectively noiseless images, we make two slightly different datasets. The first dataset (EN Set 1) used \texttt{statmorph} S/N$_{F115W}$ $> 3$ and \texttt{Flag}$_{\texttt{Galfit}}$ $=2.0$ cuts based on the effectively noiseless data, which captures fainter objects not seen in the mock CEERS dataset. The second dataset (EN Set 2) used \texttt{statmorph} S/N$_{F115W}$ $> 3$ and \texttt{Flag}$_{\texttt{Galfit}}$ $=2.0$ cuts based on the mock CEERS data, however, the effectively noiseless morphology measurements were still used as inputs to the forests. This was done to directly compare the same objects in both the CEERS-like and effectively noiseless images. Table \ref{tab:data} lists the sizes of each dataset used in this work. Note that there are fewer objects in EN Set 2 than in the mock CEERS set. Since the CEERS-like images and the nearly noiseless images are different images with different noise properties, Source Extractor will not make the exact same detections in both, and may over deblend or under deblend objects in one set of images but not the other. Additionally, \texttt{Galapagos-2} and \texttt{statmorph} may flag an object with ``bad" measurements in one set but not the other. Therefore, we are unable to perfectly match galaxies in the nearly noiseless catalog to those in the mock CEERS catalog, and lose some galaxies due to the aforementioned issues.

\subsection{Merger Definition} \label{sec:mergerdef}

We create labels for the random forest algorithm using the time since and time until a major and minor merger. Following \cite{sny2019}, we combine major and minor mergers to increase our training set size. For a binary classification scheme, galaxies that had never experienced a merger or never will experience a merger (denoted as $-1.0$ in the final catalog) were labeled as Class 0 (``non-merger"). Also following \cite{sny2019}, we choose a timescale window of 500 Myr ($\pm250$ Myr) for our merger class definition since the lifetime of merger features will likely not be longer. Therefore, galaxies that have experienced a merger greater than 250 Myr ago and will experience a merger greater than 250 Myr in the future are also assigned to Class 0. Galaxies that experienced a merger within 250 Myr, past or future, were assigned to Class 1 (``merger"). As a check, we shift the merger definition to include windows from 100 Myr to 500 Myr in the past or future (to include more pre- and post-mergers), but do not see a significant improvement in the performance of the forests. We also test using a three-class classification scheme with ``non-merger," ``past merger," and ``future merger," but the forests perform poorly in these trials, most likely due to low numbers in each merger class.

\begin{figure}
\plotone{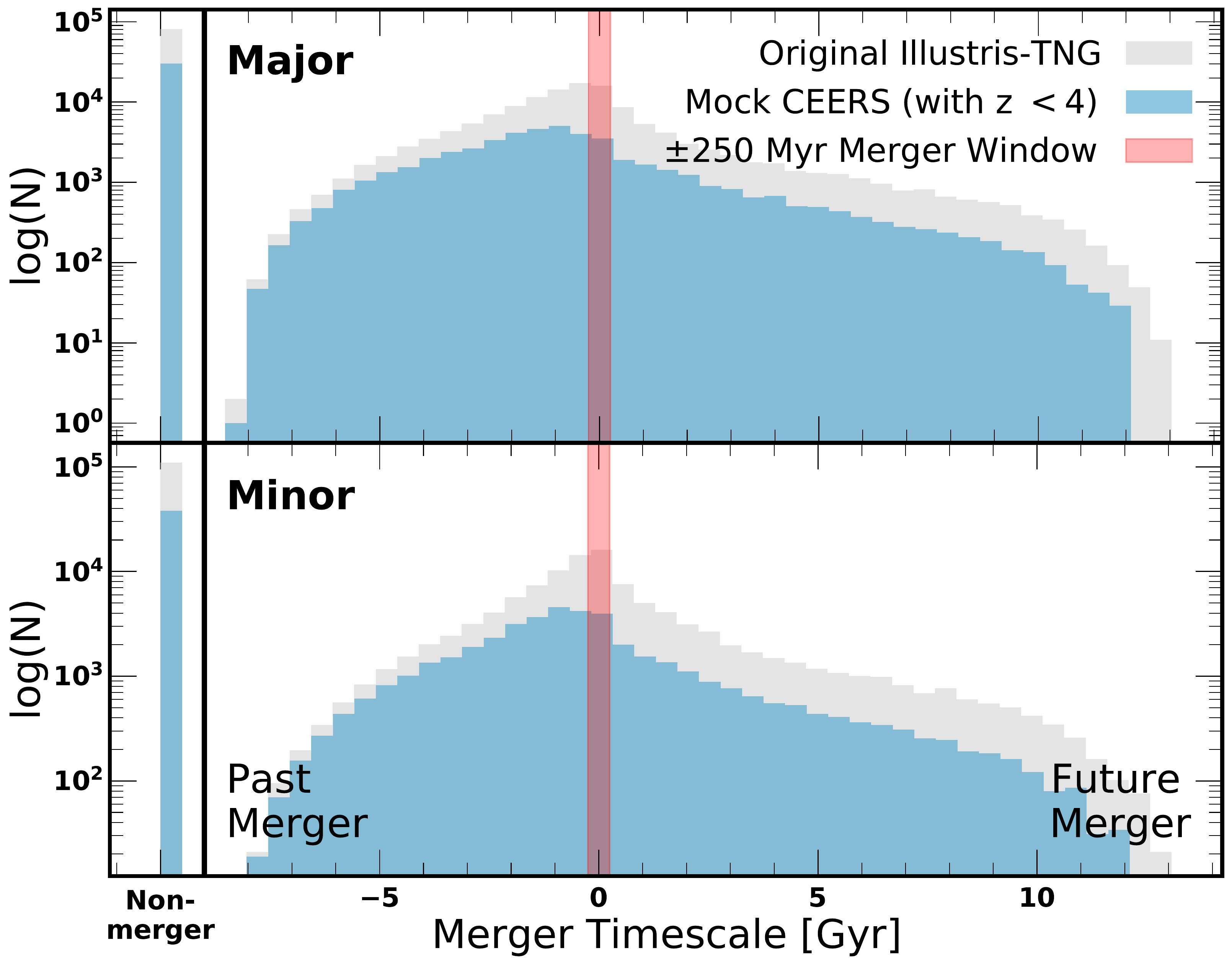}
\caption{Merger history histograms for objects from the full original IllustrisTNG sample (\textit{grey}), and from the $z < 4$ mock CEERS sample (\textit{blue}) The red shading indicates the $500$ Myr window used to distinguish between mergers and non-mergers for our random forest experiments. Each galaxy in our dataset has four merger timescales - past and future major (\textit{top}) and past and future minor (\textit{bottom}).
\label{fig:mhist}}
\end{figure}

Figure \ref{fig:mhist} shows the four merger timescales -- past major and future major (top panel) and past minor and future minor (bottom panel) -- for each galaxy in our mock CEERS set compared to the full IllustrisTNG dataset. The red shaded region shows that the selected $\pm 250$ Myr window spans a relatively narrow range of the full timescale distributions.

\subsection{Performance of Classical Methods} \label{sec:gm20}

\begin{figure}
\plotone{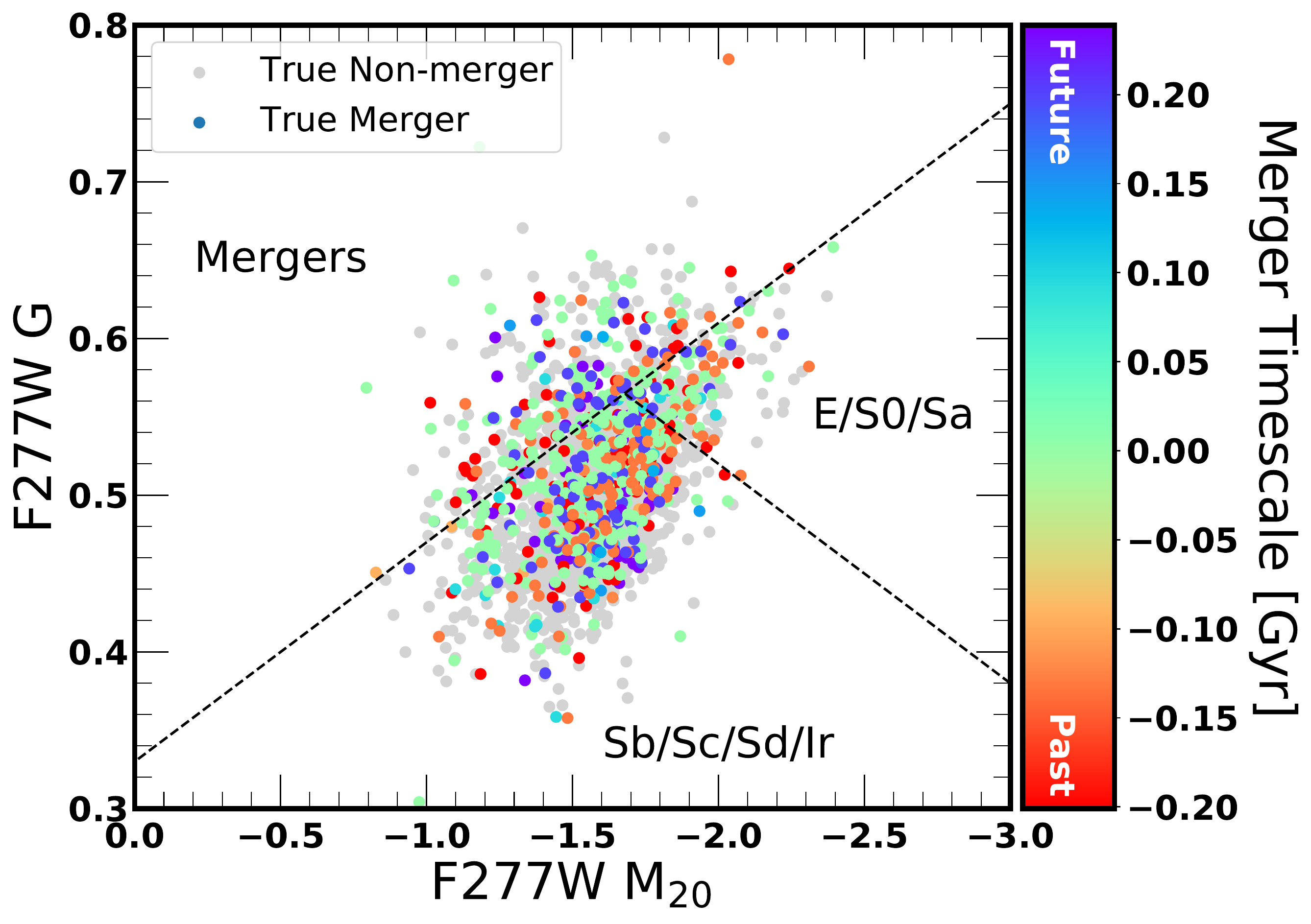}
\caption{F277W $G-M_{20}$ space for mock CEERS galaxies in the $3.0 < z < 3.5$ redshift bin. True non-mergers are \textit{grey}. True mergers are colored by the \textit{rainbow gradient} which indicates the merger timescale used for merger classifications.
\label{fig:g_v_m20}}
\end{figure}

We compare the merger classification performance of machine learning techniques with the performance of classical methods, such as the $G-M_{20}$ parameter space \citep{lotz2004, lotz2008}, in order to judge if machine learning provides any improvements. Figure \ref{fig:g_v_m20} shows the F277W observed (rest-frame optical) $G-M_{20}$ parameter space for objects in the mock CEERS $3.0 < z < 3.5$ redshift bin. The merger discriminating line is 
\begin{equation} \label{eq:gm20}
    G > -0.14 M_{20} + 0.33,
\end{equation}
as defined by \cite{lotz2008}. True mergers, according to the merger definition in \S \ref{sec:mergerdef}, occupy the same space as non-mergers. In this redshift bin, the fraction of correctly classified mergers, according to Equation \ref{eq:gm20}, is only $\sim 19\%$ since most true mergers lie below the merger discriminating line. This is to be expected since this method is very sensitive to merger stage, and is best at selecting mergers just after first passage \citep{Lotz2008b}.  Of the predicted mergers (objects above the merger discriminating line), only $\sim 48\%$ are actually true mergers. If we choose the F356W filter for our observed filter, the results are the same. Across all redshift bins, the number of correctly classified mergers ranges from $\sim 19\% - 23\%$. The number of predicted mergers that are actually true mergers ranges from only $\sim 4\%$ at the lowest redshift bin to $\sim 50\%$ at the highest redshift bin, a consequence of the increasing number of mergers at higher redshifts.

These results illustrate the poor performance of classical methods for identifying mergers at a range of stages at high redshift, and motivates the use of machine learning techniques for improving the level of completeness.

\subsection{Random Forest Experiments} \label{sec:RFexp}

The random forest (RF) algorithm is a supervised classification algorithm consisting of many decision trees \citep{ho1995,brei2001}. A single decision tree is a flowchart-like diagram where each split (or ``node") in the tree represents a decision made based on the input features. The ``terminal nodes" at the end of the tree represent the possible classifications. The algorithm uses an ensemble of decision trees to minimize overfitting from any one tree. We choose to use random forests due to their simplicity and because \cite{sny2019} demonstrated that random forests show some promise at high-z galaxy merger classification tasks.

Table \ref{tab:data} shows our dataset sizes after cleaning the dataset and defining the merger class. We split the data into training and test sets, with a training fraction of 0.67, where the ratio of objects are preserved for each class. We use \texttt{BalancedRandomForestClassifier()} from Python's \texttt{imblearn} package, which is specifically designed for imbalanced data sets. It works by deliberately undersampling the majority class during training. The morphology parameters we feed to the forests are the single S\'ersic index $n$, the two-component S\'ersic indices $n_{bulge}$ and $n_{disk}$, and the non-parametric $A, C, G, I, m_{20}, M, A_{O}, D$, and $S$, in all six filters. We also feed to the forests the non-parametric $A, C, G, I, m_{20}, M, A_{O}, D$, and $S$ as calculated from the residual images, in all filters. 

We run \texttt{keras-tuner} for hyperparameter optimization, which uses Bayesian optimization to search the parameter space and find the optimal combination of hyperparameters without having to test all possible combinations \citep{omal2019}. Generally, \texttt{keras-tuner} will find the most optimal hyperparameters in less time than other hyperparameter tuning algorithms such as \texttt{GridSearchCV}. The hyperparameters that we let vary are ``max$\_$samples" (0.9 - 0.99 with a step size of 0.1), ``max$\_$features" (1 - 15 with a step size of 1), ``max\_leaf\_nodes" (5 - 55 with a step size of 1), and ``n$\_$estimators" (1000 - 2000 with a step size of 50). We allow \texttt{keras-tuner} to search over the parameter space for 50 trials. For each trial, we provide \texttt{keras-tuner} with a cross-validation (CV) set (CV fraction was 0.2 of the training set) that preserves the ratio of objects for each class. We train seven separate forests for our seven different redshift bins.

We explore many different sets of hyperparameters and allowed ranges, and find that generally the forests perform similarly regardless of fine tuning the hyperparameters. Therefore, although further tuning is possible, we conclude that it is not necessary; the forests most likely will not perform significantly better than reported here.

We categorize the output of the random forest into four classes:
\begin{itemize}
\item True Positives (TP): the number of true mergers correctly classified by the random forest.
\item False Positives (FP): the number of non-mergers incorrectly classified as mergers.
\item True Negatives (TN): the number of correctly classified non-mergers.
\item False Negatives (FN): the number of true mergers incorrectly classified as non-mergers.
\end{itemize}
Therefore the number of RF-selected mergers is TP + FP, and the number of RF-selected non-mergers is TN + FN. The number of true mergers is TP + FN, and the number of true non-mergers is TN + FP.

We can judge the performance of the random forests using several metrics:
\begin{itemize}
\item True Positive Rate (TPR) -- also known as \textit{recall} or \textit{completeness} -- is defined as
    \begin{equation}
        TPR = \textrm{recall} = \frac{TP}{TP + FN}.
    \end{equation}
\item False Positive Rate (FPR) -- also known as \textit{fall out} - is defined as
    \begin{equation}
        FPR = \frac{FP}{FP + TN}.
    \end{equation}
\item Positive Predictive Value (PPV) -- also known as \textit{precision} -- is defined as
    \begin{equation}
        PPV = \textrm{precision} = \frac{TP}{TP + FP}.
    \end{equation}
\item F1 Score is the harmonic mean of precision (P) and recall (R), and is defined as
    \begin{equation}
    F_{1} = 2\frac{P \times R}{P + R} = \frac{TP}{TP + \frac{1}{2} (FP + FN)}.
    \end{equation}
\end{itemize}
Classifiers that perform well have both high precision and high recall (and therefore a high F1 score). Accuracy, defined as (TP + TN)/N where $N$ is the total number of objects, is a biased indicator of performance for imbalanced sets, so we do not consider it here.

\begin{figure}
\plotone{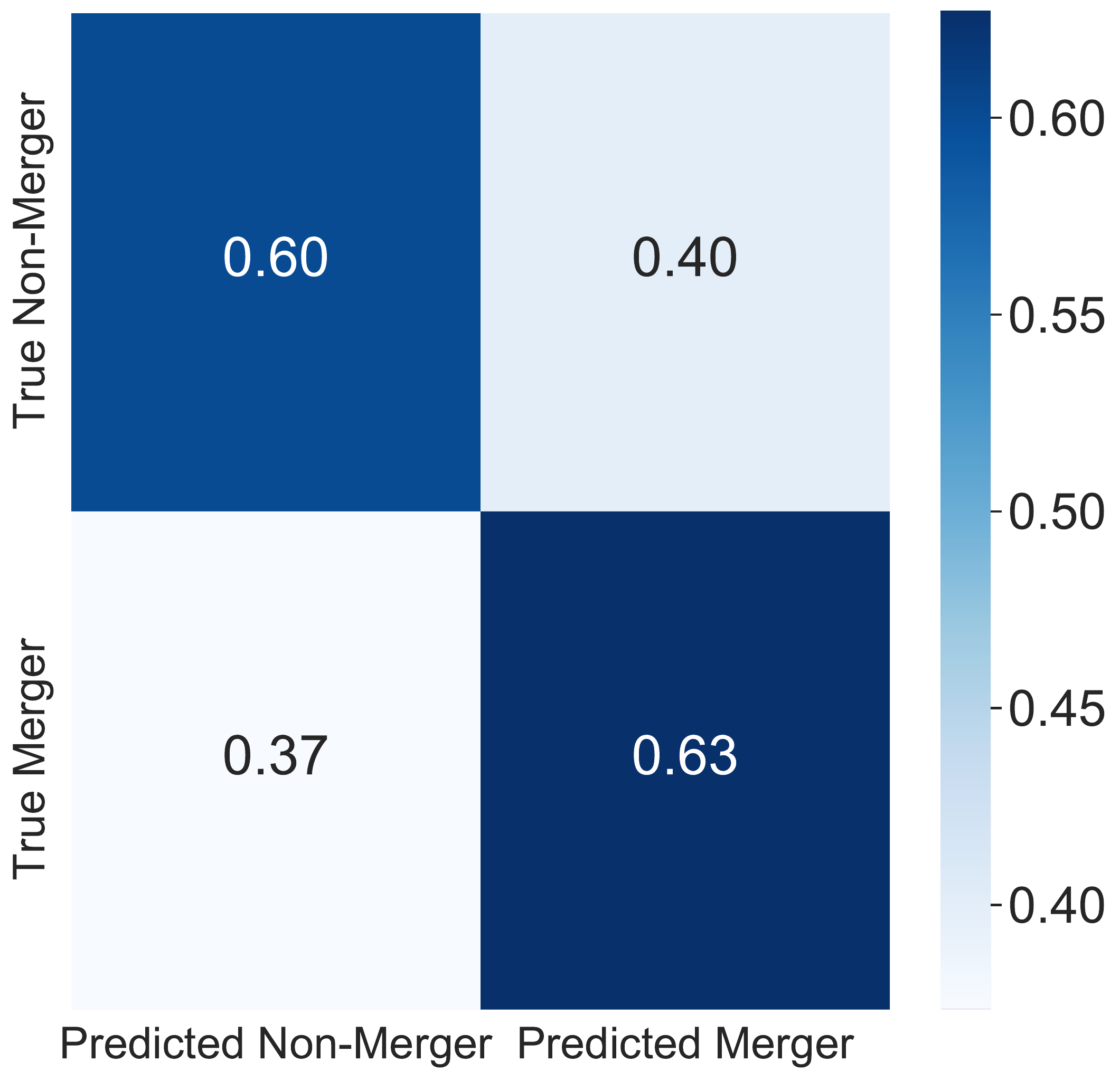}
\caption{Confusion matrix for $3 < z \leq 3.5$ objects from the test set. The diagonal shows the fraction of objects correctly classified for each class.
\label{fig:cm_3to_3p5}}
\end{figure}

\begin{figure}[t]
\centering
\minipage{0.43\textwidth}
  \includegraphics[width=1\linewidth]{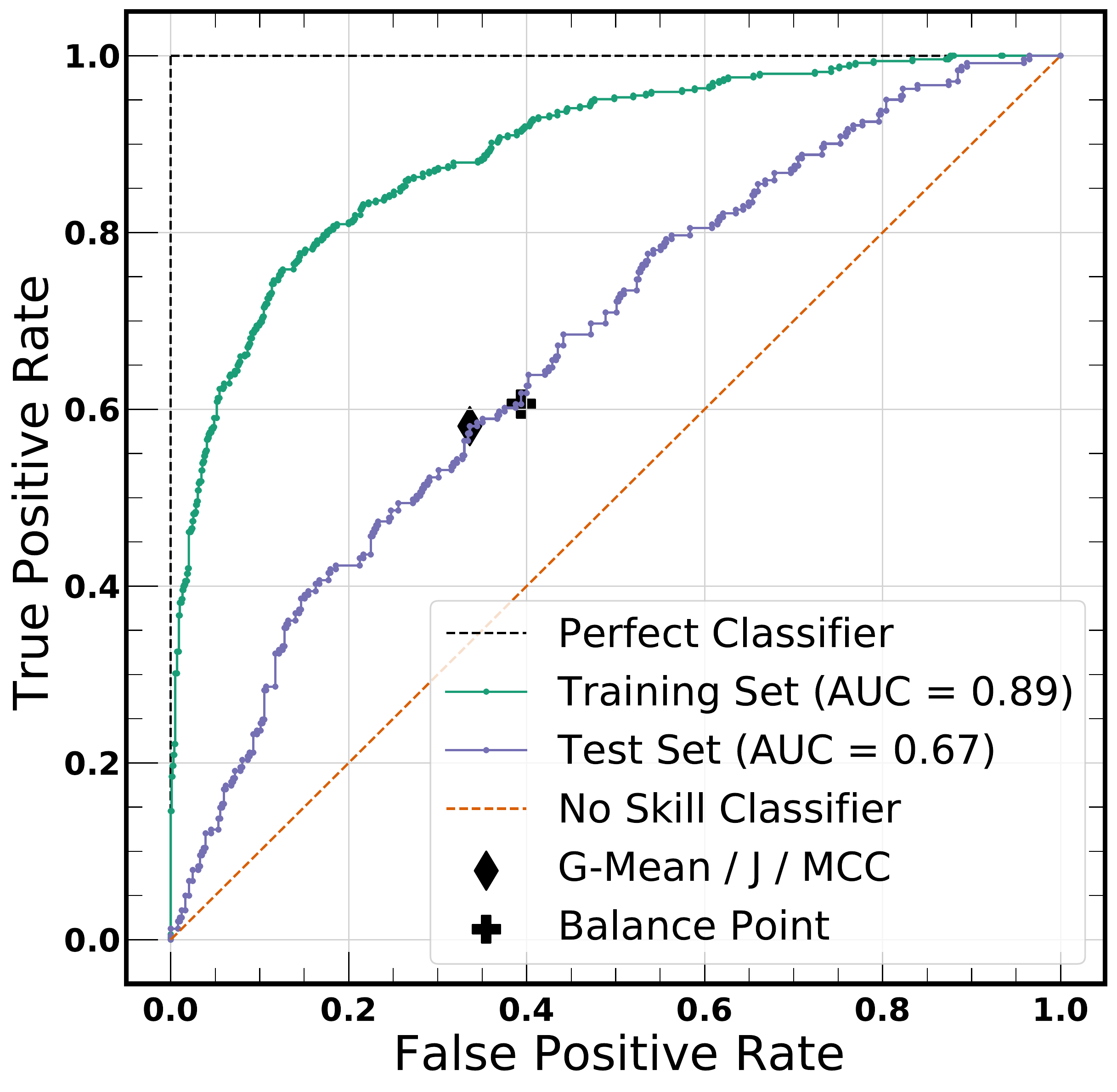}
\endminipage\hfill
\vskip2ex
\minipage{0.43\textwidth}%
  \includegraphics[width=1\linewidth]{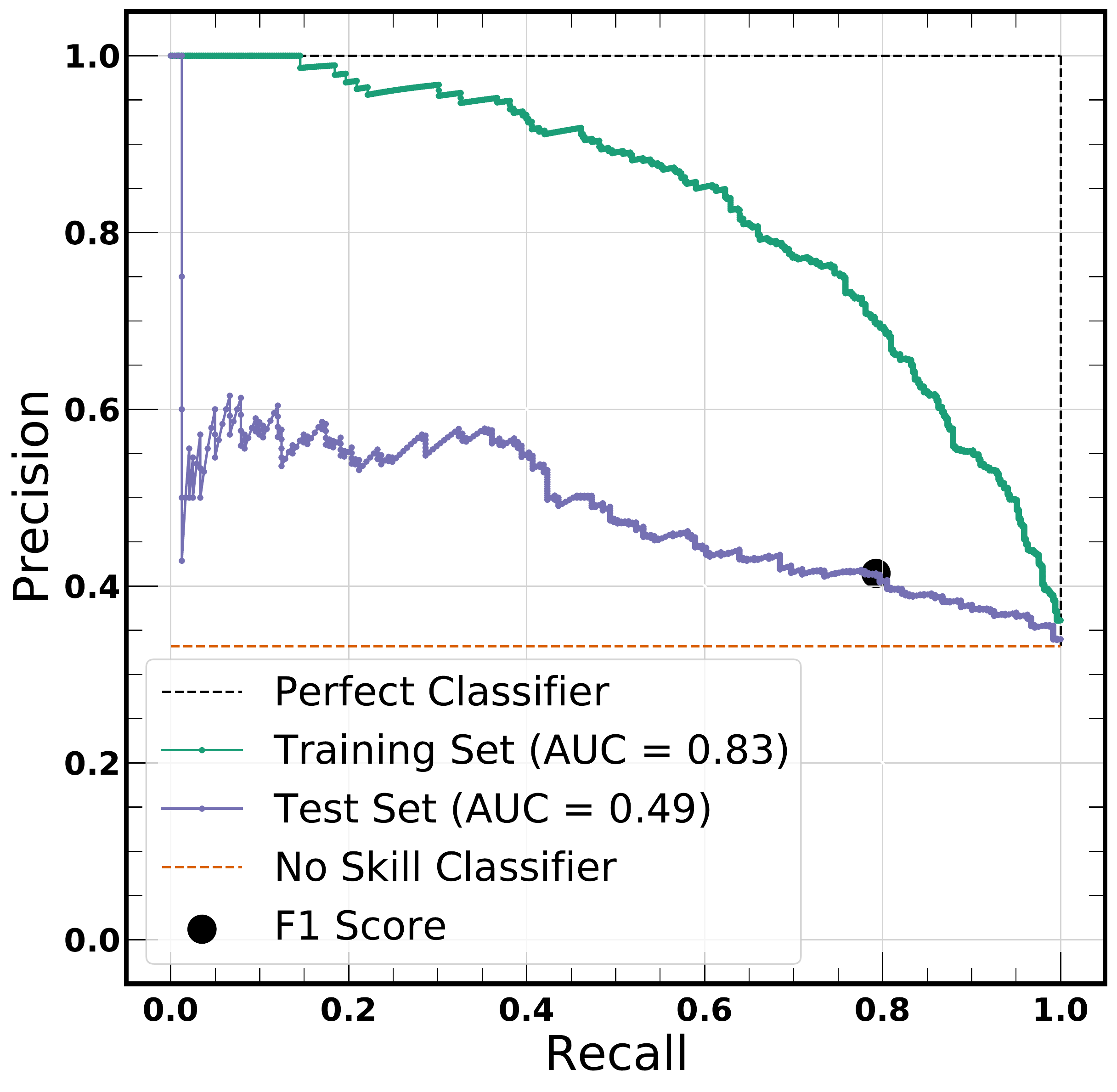}
\endminipage
\caption{\textit{\textbf{Top}}: ROC curve for the $3 < z \leq 3.5$ random forest. The \textit{black diamond} shows the optimal threshold selected by G-Means, J statistic, and MCC and the \textit{black cross} shows the optimal threshold selected by the balance point (see \S \ref{sec:thresh}). \textit{\textbf{Bottom}}: Precision-recall curve for the $3 < z \leq 3.5$ random forest. The \textit{black circle} shows the optimal threshold selected by the f1 score (see \S \ref{sec:thresh}). In both plots, the training set (\textit{green}) and test set (\textit{purple}) curves lie in the region of ``good'' classifiers (between the perfect (\textit{black}) and no skill (\textit{red}) classifiers). 
\label{fig:roc_pr_curve}}
\end{figure}

\begin{figure}
\epsscale{1.2}
\plotone{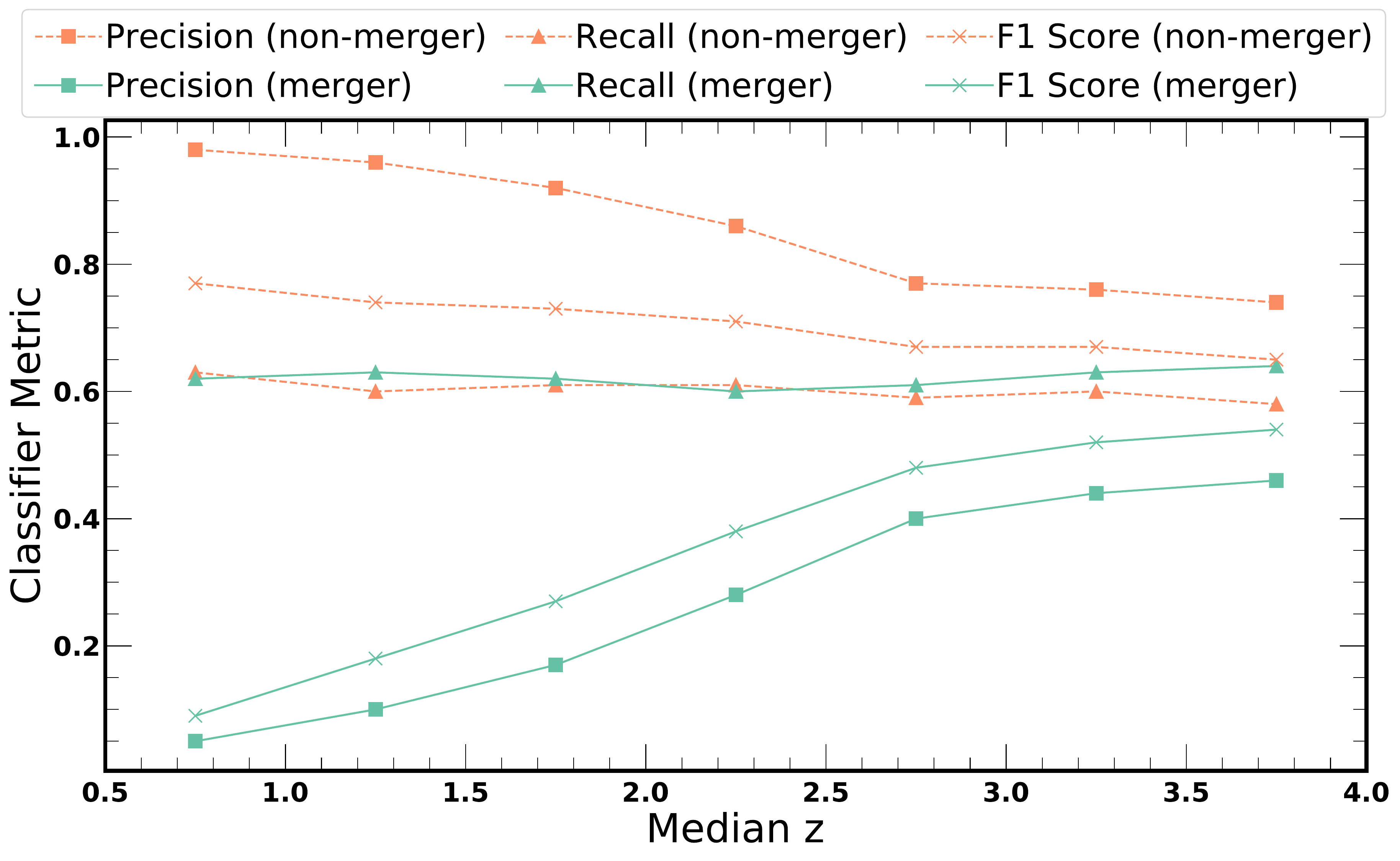}
\caption{Precision (\textit{squares}), recall (\textit{triangles}), and f1 score (\textit{crosses}) for mergers (\textit{green}) and non-mergers (\textit{dashed orange}) as a function of redshift. Non-merger class metrics tend to worsen with redshift while merger class metrics tend to improve with redshift. The exception is recall, which is consistent around $\sim 0.60$ across redshift.
\label{fig:metrics_comp}}
\end{figure}

Figure \ref{fig:cm_3to_3p5} shows a confusion matrix for $3 < z \leq 3.5$ galaxies. This is an example from one of the best random forest trials. This shows that the forest correctly classified $60\%$ of the non-merger class and $63\%$ of the merger class. The confusion matrices for the other redshift bins all look similar to this one, where $58\% - 63\%$ of the non-merger class were correctly classified and $60\% - 64\%$ of the merger class were correctly classified (see the recall values in Figure \ref{fig:metrics_comp}).

The top panel of Figure \ref{fig:roc_pr_curve} shows the corresponding receiver operating characteristic (ROC) curve for this redshift bin, which illustrates the performance of the random forest for different discrimination thresholds. A discrimination threshold is the probability cutoff (default $= 0.5$) used to assign the final classification. The curve of a perfect classifier would consist of two straight lines from (0,0) to (0,1) and from (0,1) to (1,1). The curve of a random classifier consists of a straight line from (0,0) to (1,1). The performance of classifiers can then be judged by how close they are to the upper left corner of the plot. This figure shows that our random forest does better than a random classifier. 

The bottom panel of Figure \ref{fig:roc_pr_curve} shows the corresponding precision-recall curve for this redshift bin, which in principle is more informative for imbalanced datasets than the ROC curve. For the precision-recall curve, a perfect classifier reaches the upper right-hand corner of the plot at (1,1). The curve of a random classifier is not fixed, like in the ROC curve, but determined by the ratio of positives (mergers) to total number of objects. Therefore a random classifier for a perfectly balanced dataset would lie at $y=0.5$. This plot, like the ROC curve, also shows that our forest performs better than random chance. However these plots, especially the precision-recall curve, also show that our classifiers are far from perfect. The ROC curves and precision-recall curves for the other redshift bins look similar to those shown here.

\begin{figure*}[t]
\plottwo{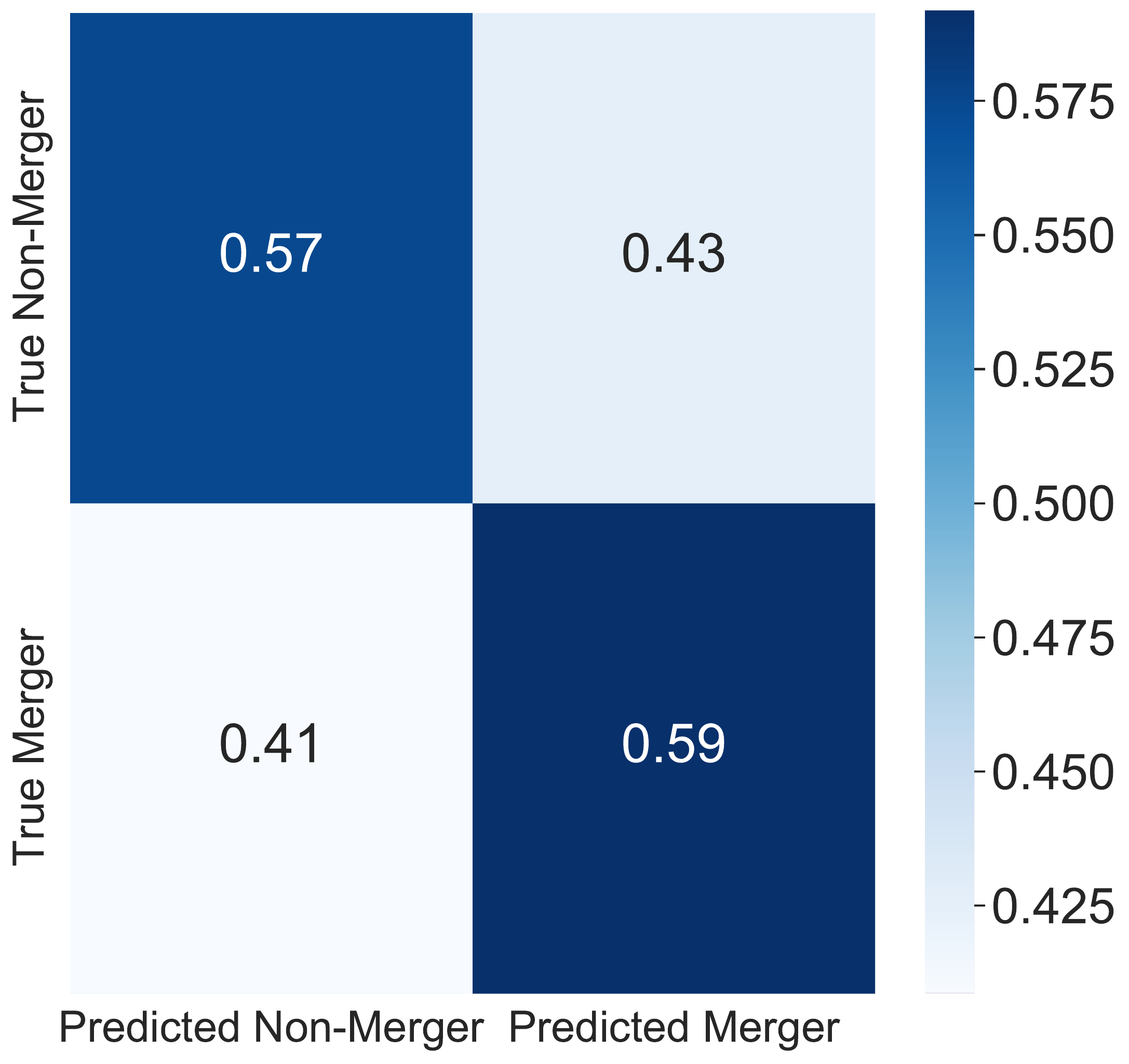}{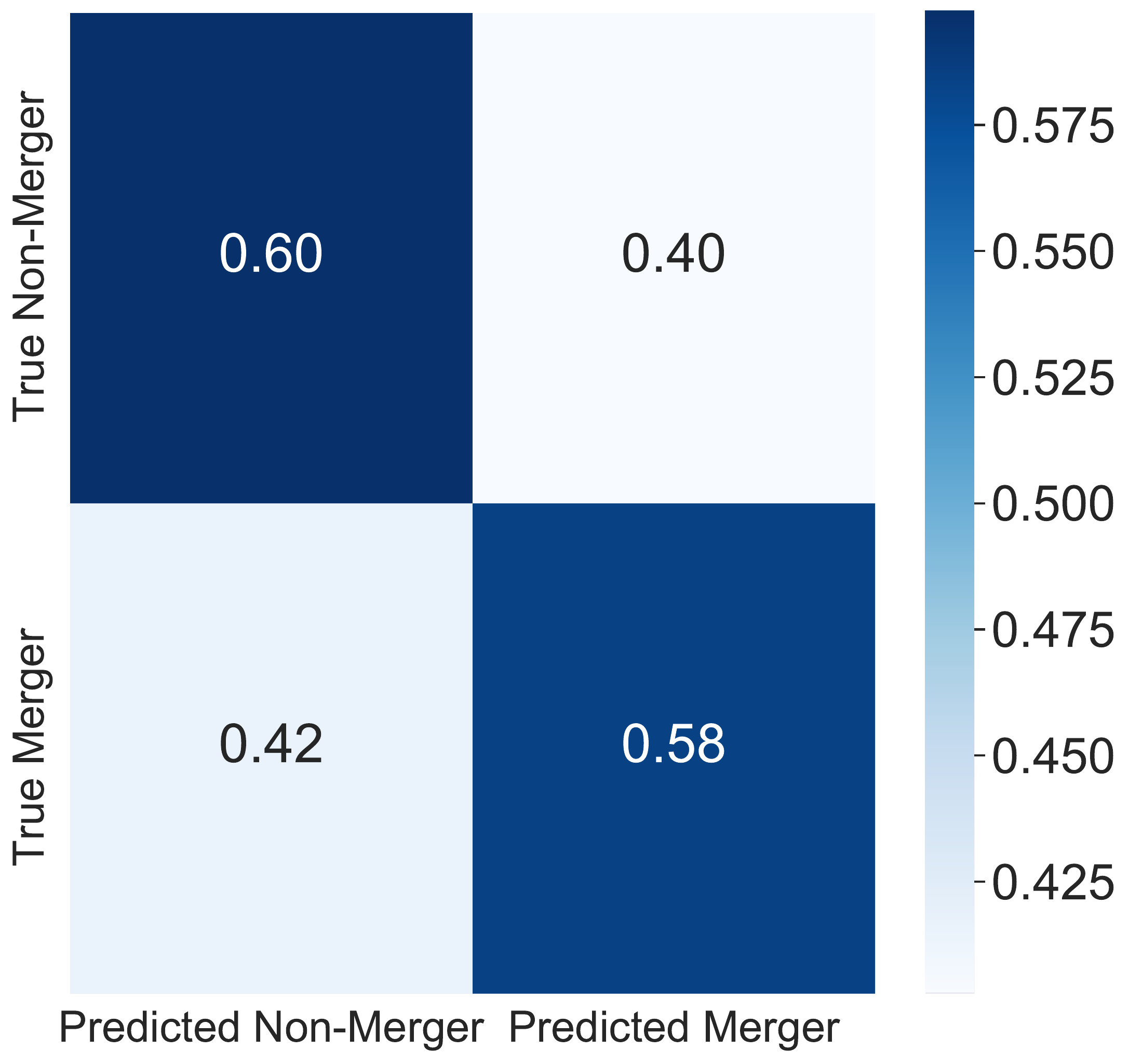}
\caption{Confusion matrices for $3.0 < z \leq 3.5$ objects from the EN Set 1 (\textit{\textbf{left}}) and from the EN Set 2 (\textit{\textbf{right}}).
\label{fig:ensets}}
\end{figure*}

Figure \ref{fig:metrics_comp} shows how precision, recall, and f1 scores change across redshift for both mergers and non-mergers. For mergers, the classification metrics improve as redshift increases. For non-mergers, the classification metrics generally worsen as redshift increases. This is likely due to the test set becoming more balanced at higher redshifts. To test this, we artificially balanced the $z = 0.5 - 1.0$ test set and found that the merger precision score (and therefore f1 score) dramatically increased and the non-merger precision (and therefore f1 score) decreased such that they were more in line with the results of the later redshift bins. This means that the performance of the $z = 0.5 - 1.0$ forest can be improved with respect to the merger class by simply randomly removing non-mergers from the test set. This implies that the better performance of the forests of the later redshift bins is mostly due to the increasing lack of non-mergers, not because the forest was better trained.

\begin{figure*}[t]
\plotone{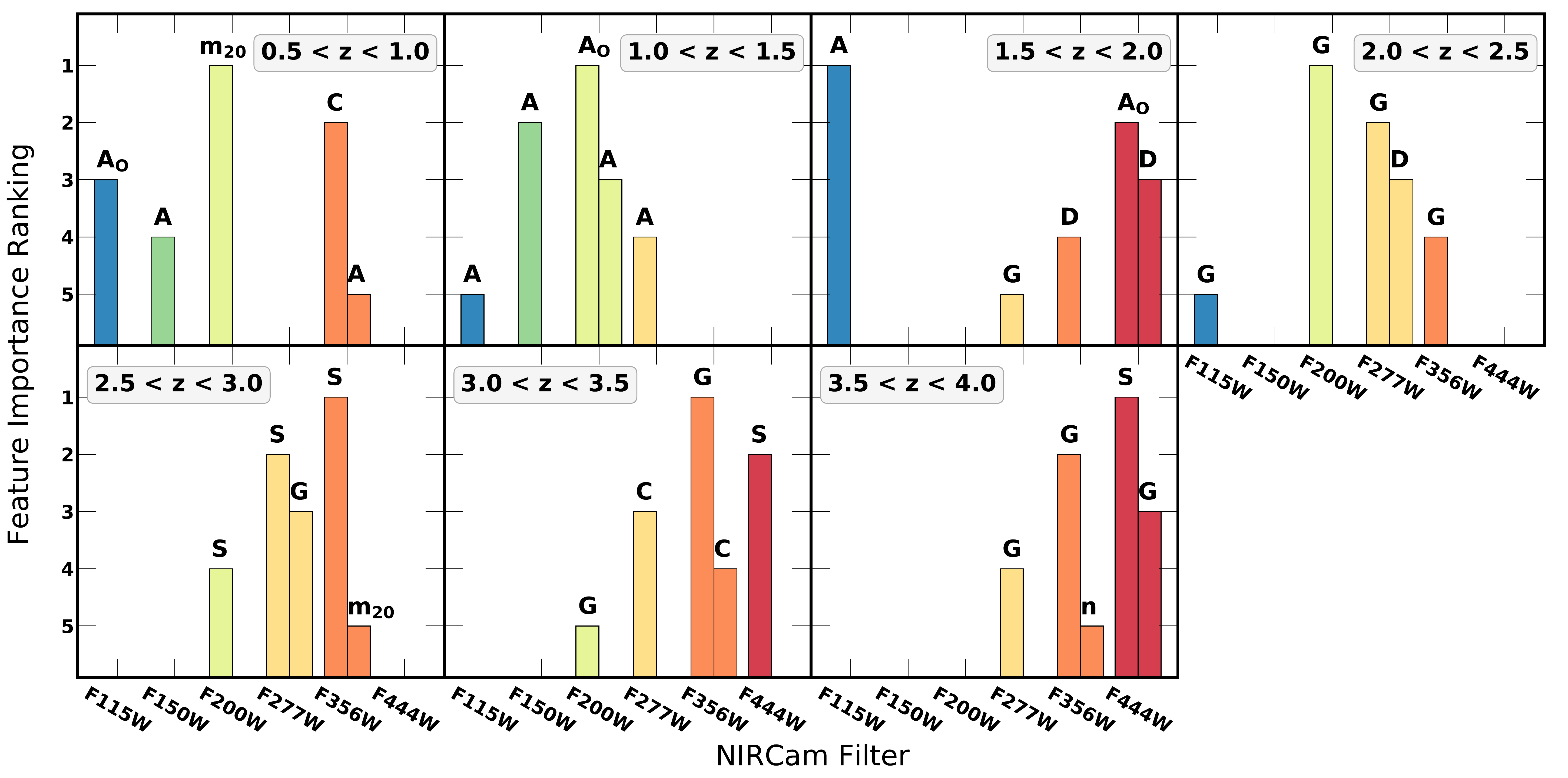}
\caption{The top five most important features (1 - most important, 5 - 5\textsuperscript{th} most important) in each redshift bin, as a function of filter. Bluer colors correspond to bluer filters, and redder colors correspond to redder filters. The abbreviations are: $A_{O}$ - outer asymmetry, $A$ - asymmetry, $C$ - concentration, $D$ - deviation, $G$ - Gini statistic, $m_{20}$ - moment of light, $M$ - multimode, $n$ - S\'ersic index, $S$ - clumpiness. See \S \ref{sec:params} and references therein. Asymmetry features in the bluer filters are more important for low redshift bins while bulge/clump features in the redder filters are more important for higher redshift bins.
\label{fig:feat_import}}
\end{figure*}

For the effectively noiseless images, the forests trained and tested on EN Set 1 generally performed worse than the mock CEERS forests. The train and test sets were larger, but this increase was mostly due to the inclusion of faint and probably ambiguous-looking galaxies that were cut from the mock CEERS set. We conclude that the difficulty of classifying these objects probably out-weighed any gains from the ability to see fainter structures in the effectively noiseless images. The left panel of Figure \ref{fig:ensets} shows the confusion matrix for this set.

The forests trained and tested on EN Set 2 generally performed slightly worse than the mock CEERS forests. In this case, any gains from the effectively noiseless images are probably out-weighted by the smaller size of the training and test sets. The right panel of Figure \ref{fig:ensets} shows the confusion matrix for this set.

\section{Discussion} \label{sec:discuss}

\subsection{Feature Importance}

Each forest calculates the importance of the features given to it. The more important a feature is, the more useful it is to the forest for determining the difference between mergers and non-mergers. Figure \ref{fig:feat_import} shows the top five most important features for each redshift bin. This figure shows that asymmetry features (e.g., $A$ and $A_{O}$) are most important for low redshift bins while bulge and clump features (e.g., $G$ and $S$) are more important for higher redshift bins. These most important features are calculated from the science images, not the residual images. There also appears to be a dependence on filter. The bluer F115W filter is more useful for low redshift bins, and the redder F444W filter is more useful for higher redshift bins. This seems to indicate that the forests are using the rest-frame optical features to make decisions, even though all filters were available to each forest.

\begin{figure*}
\includegraphics[scale=0.6]{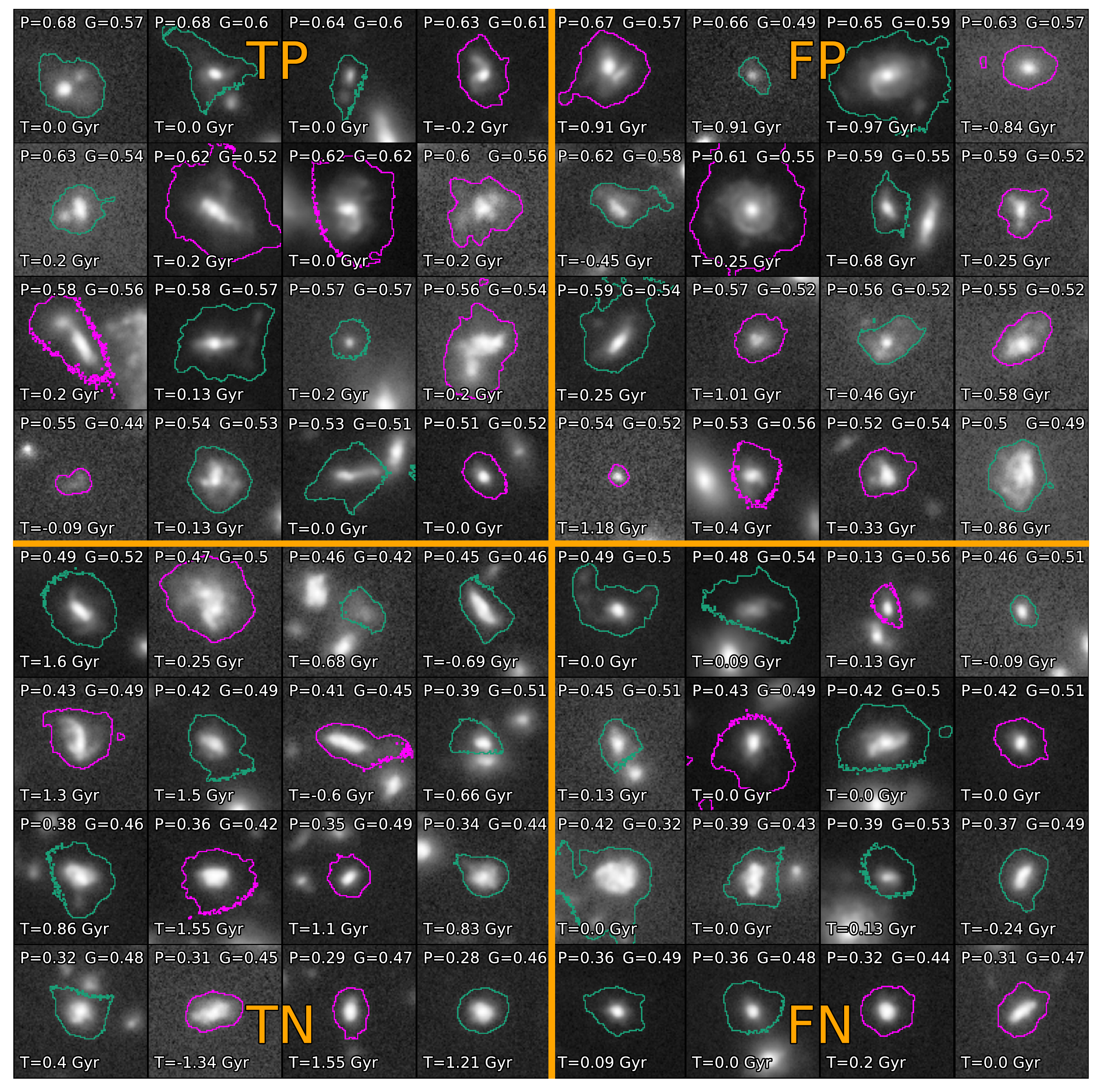}
\caption{Examples of true positive (TP), false positive (FP), true negative (TN), and false negative (FP) galaxies in the F356W filter from the $3.0 < z < 3.5$ redshift bin. Each stamp is 3 x 3 arcsec. In each stamp: the merger probability output by the forest is in the upper left, the F356W Gini statistic is in the upper right, and the timescale since or until the most recent merger (major or minor) is in the bottom left. The outlines show the segmentation map, color-coded by major (\textit{magenta}) and minor (\textit{green}) mergers, respectively.
\label{fig:example_grid}}
\end{figure*}

Figure \ref{fig:example_grid} shows examples of $3.0 < z < 3.5$ galaxies categorized into true positives (correctly classified mergers), false positives (incorrectly classified non-mergers), true negatives (correctly classified non-mergers), and false negatives (incorrectly classified mergers). The top left hand corner of each stamp shows the probability of the object being a merger as assigned by the random forest. The stamps are arranged in order of decreasing probability, so the horizontal orange line effectively represents the probability threshold between merger and non-merger classifications, which is the default 0.5. The top right hand corner of each stamp shows the F356W Gini statistic for each galaxy, which was the most important feature for the random forest. The bottom left hand corner of each stamp shows the merger timescale for each galaxy. The timescales include both major and minor mergers. A positive timescale indicates the time since a past merger, while a negative timescale indicates the time until a future merger. Galaxies can have both past and future mergers, so whichever timescale is smallest is shown here. Recall that the merger cutoff defined in \S \ref{sec:mergerdef} is $\pm250$ Myr, so true positives and false negatives all have a merger timescale $< 0.25$ Gyr, while false positives and true negatives all have a merger timescale $> 0.25$ Gyr. Finally, the segmentation map outlines are color-coded by merger type (major or minor).

The distribution of merger probabilities ranges from about 0.3 - 0.7, while the distribution of the Gini statistic ranges from about 0.4 to 0.65. There is a slight trend where probability increases as Gini increases, which becomes increasingly clear when examining the full test set. This is expected since the F356W Gini statistic was the most important feature to the random forest for this redshift bin, but the correlation is not so strong that Gini can solely be used to determine merger status. There appears to be little to no trend of merger timescale with either Gini or with probability when looking at the full test set.

Other interesting insights come from looking at the false negatives and false positives. Many of the false positives in Figure \ref{fig:example_grid} have segmentation maps that appear elongated, either because the segmentation map is contaminated with emission from background or neighboring galaxies, or because the galaxy has persisting remnants of signatures from a merger outside of the chosen time frame, which may have contributed to the ``merger" designation by the forest. The average past and future timescales of true negatives is $0.75 \pm 0.35$ Gyr and $0.66 \pm 0.26$ Gyr, respectively. The average past and future timescales of false positives is $0.66 \pm 0.32$ Gyr and $0.71 \pm 0.43$ Gyr, respectively. Since the timescale distribution of false positives generally matches that of the true negatives, within error, it does not appear that false positives are more likely to be closer in time to having a merger than other non-mergers. On the other hand, many of the false negatives appear relatively undisturbed visually, especially the ones with smaller merger probabilities, suggesting that even though these are true mergers those mergers have had a relatively minor impact on the morphology. The fraction of minor mergers among the true positives and false negatives is $35.1\%$ and $35.6\%$, respectively. This indicates that false negatives are no more likely to be minor mergers than true positives.

\subsection{Thresholding} \label{sec:thresh}

\begin{figure}
\centering
\minipage{0.383\textwidth}
  \includegraphics[width=\linewidth]{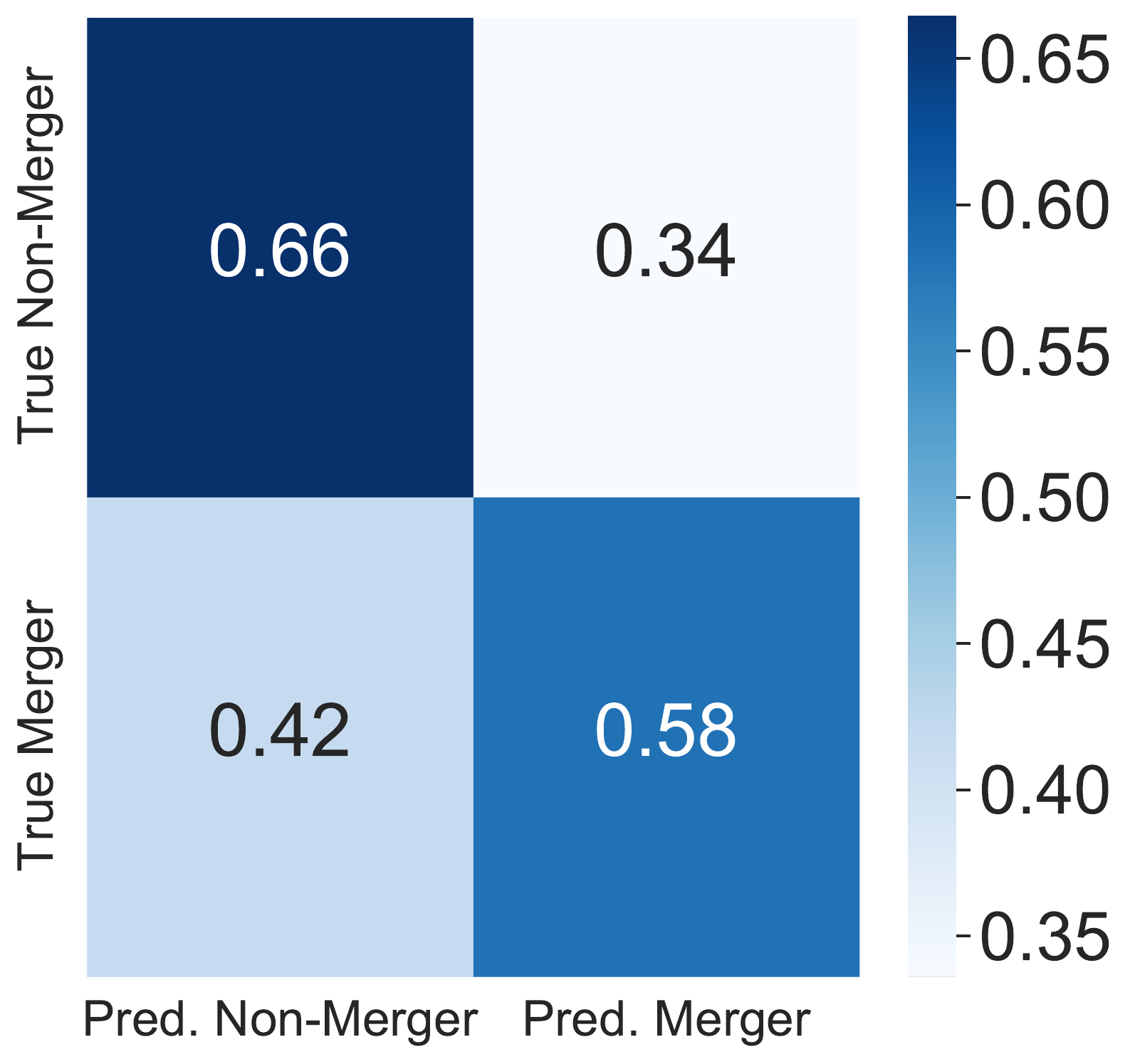}
\endminipage\hfill
\vskip2ex
\minipage{0.383\textwidth}%
  \includegraphics[width=\linewidth]{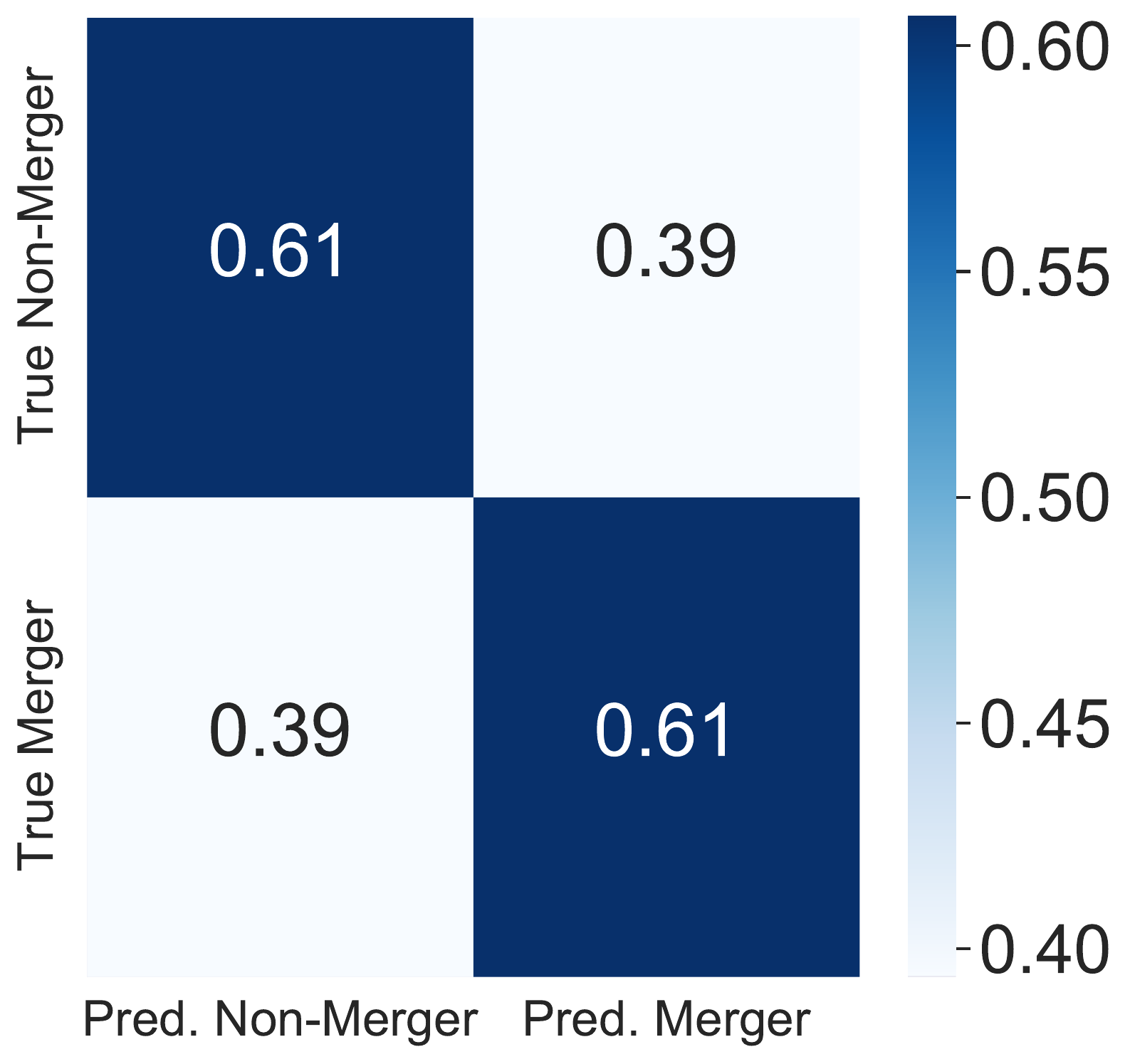}
\endminipage\hfill
\vskip2ex
\minipage{0.383\textwidth}%
  \includegraphics[width=\linewidth]{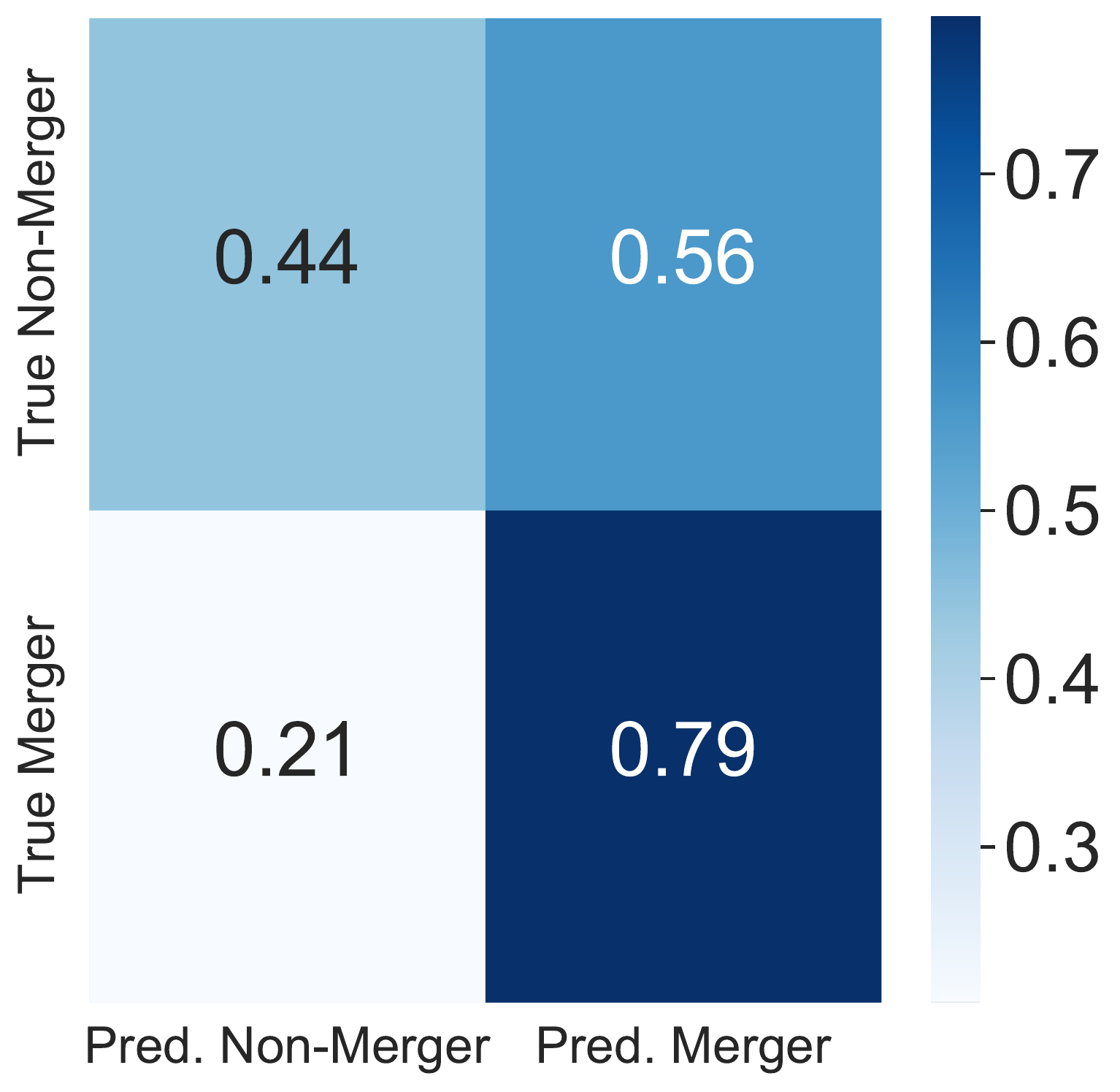}
\endminipage
\caption{Confusion matrices for $3.0 < z \leq 3.5$ objects from the mock CEERS set using the G-Mean threshold (\textit{\textbf{top}}), the balance point threshold (\textit{\textbf{middle}}), and the F1 score threshold (\textit{\textbf{bottom}}).} \label{fig:threshs}
\end{figure}

The probability threshold used to distinguish between mergers and non-mergers can be adjusted in an attempt to improve the performance of the random forest. There are a number of methods for selecting the optimal threshold based on the trade-off between the TPR and FPR, or precision and recall, including:
\begin{itemize}
\item Geometric Mean (G-Mean), defined by 
    \begin{equation}
    \textrm{G-Mean} = \sqrt{TPR \times (1 - FPR)}.
    \end{equation}
\item Youden's J Statistic \citep{you1950}, defined by
    \begin{equation}
    \textrm{J} = TPR - FPR.
    \end{equation}
\item Matthew's Correlation Coefficient (MCC) \citep{mat1975}, defined by 
    \begin{multline}
    \textrm{MCC} = \\ 
    \frac{TP \times TN - FP \times FN}{\sqrt{(TP + FP)(TP + FN)(TN + FP)(TN+FN)}}.    
    \end{multline}
\item Balance Point, defined by the point at which $TPR = 1 - FPR$.
\item F1 Score - defined in \S \ref{sec:RFexp} - which is the harmonic mean between precision and recall.
\end{itemize}

For the $3.0 < z < 3.5$ redshift bin, G-mean, J, and MCC all return the same optimal threshold of 0.518. The balance point returns a threshold of 0.504. All four are very close to the default threshold of 0.5. Figure \ref{fig:roc_pr_curve} highlights the location of these two thresholds on the ROC curve, which are located where the curve of the test set is closest to the upper left hand corner of the plot. The optimal threshold based on the F1 score was 0.470, which is highlighted in the precision-recall curve in Figure \ref{fig:roc_pr_curve}. Here, the threshold is located where the curve of the test set is closest to the upper right hand corner of the plot.

Use of the G-mean/J/MCC threshold (Figure \ref{fig:threshs}, \textit{top}) or the balance point threshold (Figure \ref{fig:threshs}, \textit{middle}) improves the performance of the forest on the non-merger class (fraction correctly classified = 0.66 or 0.61, respectively, where before it was 0.60), at a cost of poorer performance on the merger class. Use of the F1 score threshold (Figure \ref{fig:threshs}, \textit{bottom}) drastically improves the performance on the merger class (fraction correctly classified = 0.79 where before it was 0.63), but at great cost to the non-merger class (fraction correctly classified is 0.44 where before it was 0.60). This shows one could optimize the performance of the forest to generate a more complete sample of mergers, but that sample would be highly contaminated. None of the thresholds improve performance on both the merger and non-merger classes.

\subsection{Merger Fraction and Merger Rate}

\begin{figure*}[t]
\plottwo{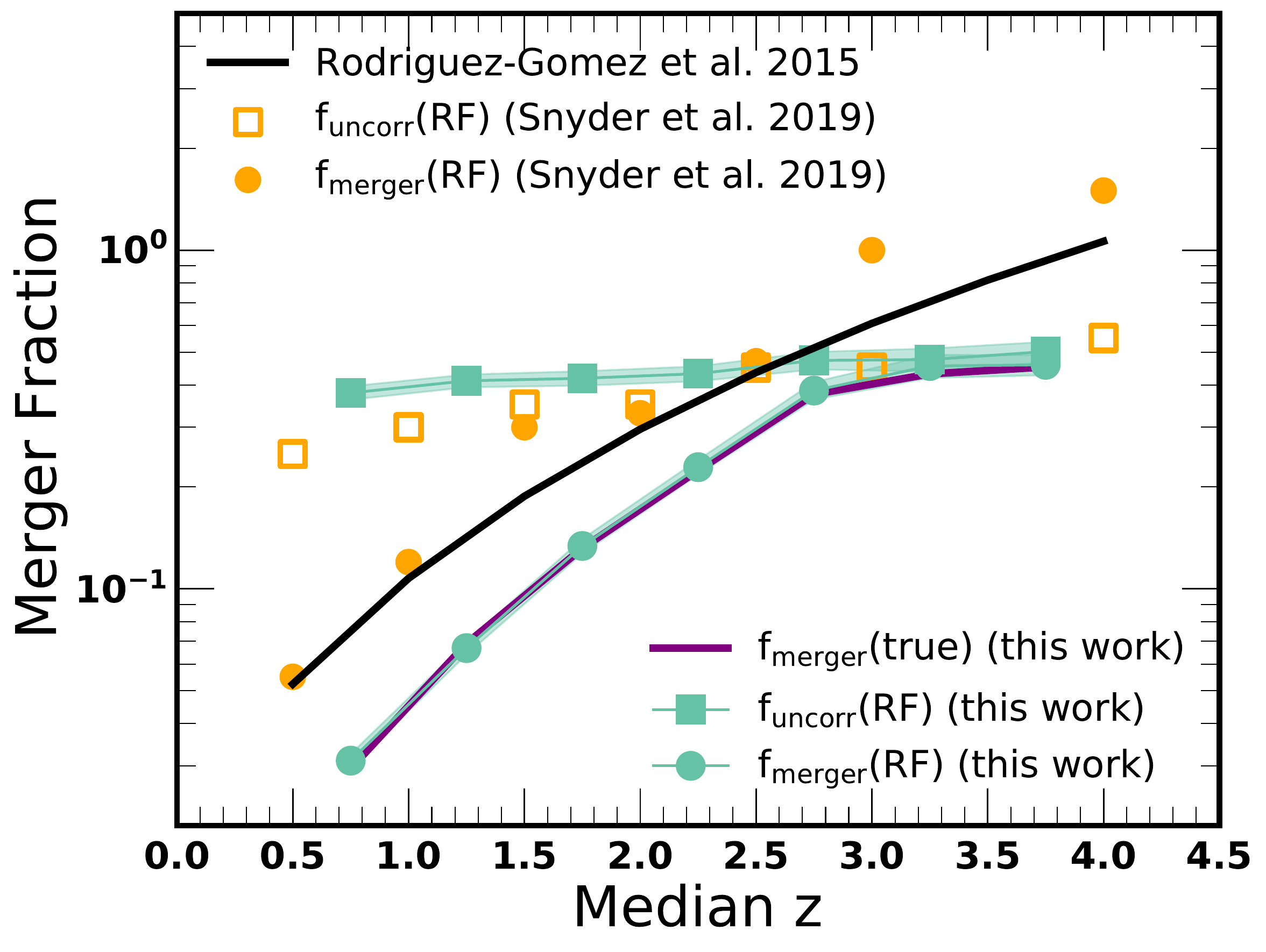}{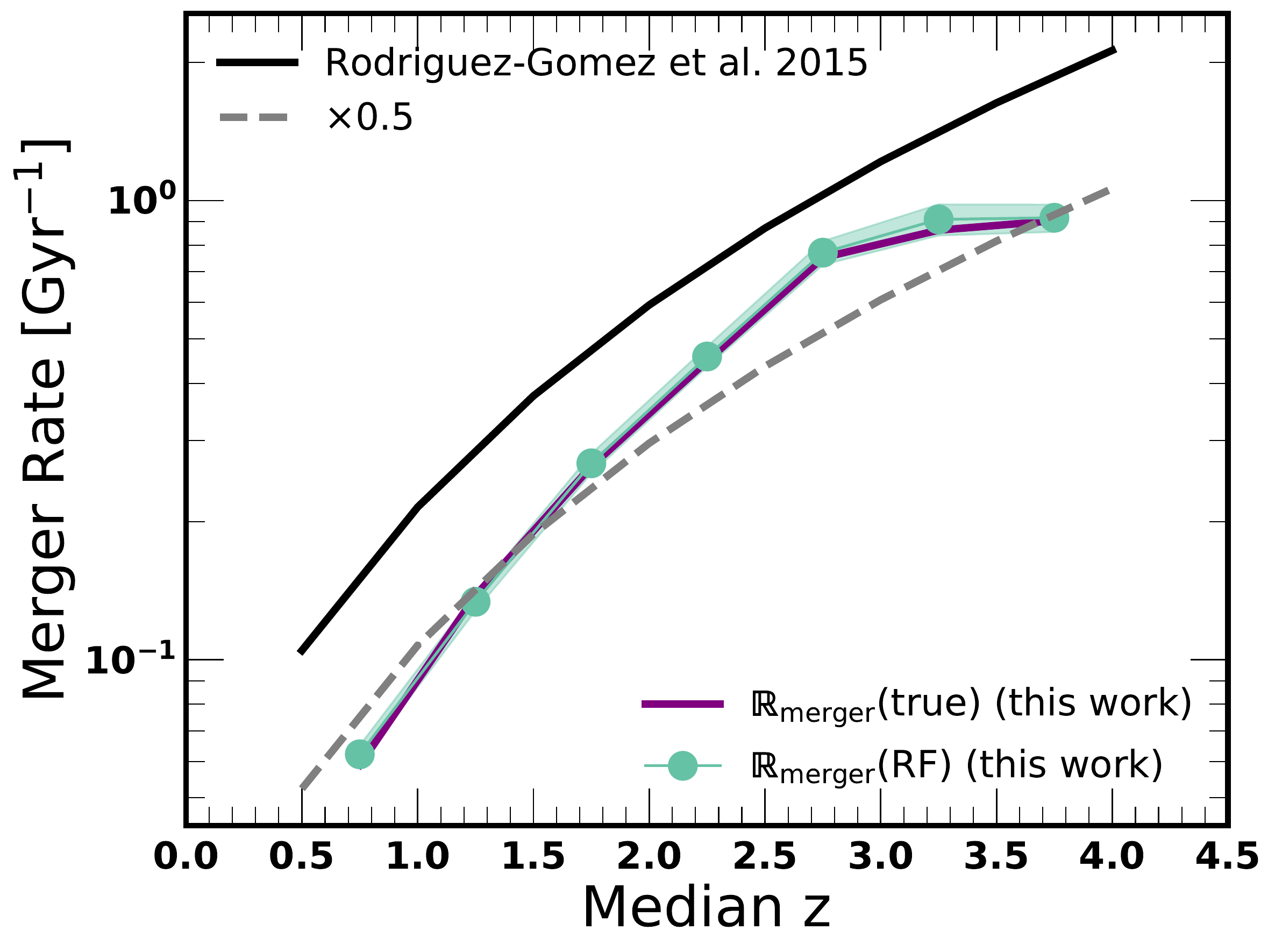}
\caption{\textit{\textbf{Left}}: This work's uncorrected RF-selected merger fractions $f_{\mathrm{uncorr}}$(RF) (\textit{green squares}), corrected RF-selected merger fractions $f_{\mathrm{merger}})$(RF) (\textit{green circles}), and the true merger fraction $f_{\mathrm{merger}}$(true) (\textit{purple line}) as compared to the uncorrected (\textit{orange open squares}) and corrected (\textit{orange circles}) RF-selected merger fractions from \cite{sny2019}, as well as the theoretical Illustris merger fraction (\textit{black line}) derived from \cite{rod2015}. The green shaded regions indicate the binomial $95\%$ confidence interval \citep{cam2011} for this work's $f_{\mathrm{uncorr}}$(RF) and $f_{\mathrm{merger}}$(RF). \textit{\textbf{Right}}: This work's RF-selected merger rate $\mathbb{R}_{\mathrm{merger}}$(RF) (\textit{green circles}) and true merger rate $\mathbb{R}_{\mathrm{merger}}$(true) (\textit{purple line}) as compared to the theoretical Illustris merger rate (\textit{black line}) derived from \cite{rod2015}. The green shaded regions again indicate the binomial $95\%$ confidence interval for this work's $\mathbb{R}_{\mathrm{merger}}$(RF).
\label{fig:merg_frac_rate}}
\end{figure*}

Finally, we calculate the merger fraction and merger rate using both mergers selected by the random forest and true mergers based on our merger timescale window of 0.5 Gyr. First, we calculate the fraction of merging galaxies selected by the random forest as $f_{\mathrm{uncorr}}(\mathrm{RF}) = N_{RF}/N$, that is, the total number of galaxies (in the test set) selected as mergers by the random forest divided by the total number of galaxies (in the test set) for a given redshift bin. Then we multiply this fraction by $PPV / TPR$ and $<M/N>$ \citep{sny2019}. $PPV / TPR$ corrects for the known incompleteness and purity of the classifier based on the training set. $<M/N>$ is the average number of merging events per true merger, and accounts for the fact that some true mergers experience more than one merger during the specified time frame of 0.5 Gyr. Therefore, the actual merger fraction for the RF-selected mergers is:

\begin{equation}
    f_{\mathrm{merger}}(\mathrm{RF}) = \frac{N_{RF}}{N} \frac{PPV}{TPR} <\frac{M}{N}> .
\end{equation}
Here, $<M/N>$ was calculated using only the true positives in the test set.

The merger fraction for the true merging galaxies (from the test set) based on our merger timescale definition is then

\begin{equation}
    f_{\mathrm{merger}}(\mathrm{true}) = \frac{N_{true}}{N} <\frac{M}{N}>,
\end{equation}
since there is no need to correct for the performance of the random forest classifier. But our intrinsic merger definition also does not account for multiple mergers with our time frame. Here, $<M/N>$ was calculated using the true positives plus the false negatives in the test set.

The left panel of Figure \ref{fig:merg_frac_rate} shows our uncorrected random forest merger fraction $f_{\mathrm{RF}}$, the corrected random forest merger fraction $f_{\mathrm{merger}}$(RF), and the true merger fraction $f_{\mathrm{merger}}$(true). This plot reveals that the uncorrected fraction $f_{\mathrm{RF}}$ is overestimated compared to the true merger fraction, but once corrected by $PPV / TPR$, the RF-selected merger fractions and the true merger fractions line up very well. This panel also shows the theoretical Illustris merger fraction \cite[derived from][]{rod2015} and the random forest merger fraction estimated from \cite{sny2019}. Our uncorrected merger fractions line up very closely with the uncorrected merger fractions from \cite{sny2019}. The shape and steep slope of our corrected fraction $f_{\mathrm{merger}}$(RF) and true fraction $f_{\mathrm{merger}}$(true) generally match the theoretical Illustris fraction. However, our $f_{\mathrm{merger}}$(RF) and $f_{\mathrm{merger}}$(true) are underestimated compared to theory and the corrected random forest fractions from \cite{sny2019}.

To calculate merger rates, we divide the merger fractions by our merger window timescale of $0.5$ Gyr. The right panel of Figure \ref{fig:merg_frac_rate} shows our merger rates for the corrected RF-selected mergers and for true mergers. We again show the theoretical Illustris merger rates derived from \cite{rod2015} as shown in \cite{sny2019}. This panel shows that our merger rates are underestimated by a factor of about 0.5 (\textit{grey dashed line}) when compared to the theoretical merger fraction.

Since both our random forest fractions and rates, and true fractions and rates are underestimated when compared to theory, the issue may lie in our calculation of $<M/N>$. From our merger history catalog, we calculate the timescales for the \textit{most recent} and the \textit{first future} major and minor mergers. However, this does not tell us if a galaxy has experienced e.g., more than one past major merger or more than one future minor merger within the specified merger window. Therefore our values for $<M/N>$ are almost certainly underestimated, which may account for the discrepancy between our data and the theoretical Illustris curves.

\subsection{Comparison to Previous Works}

We compare our results to that of \cite{sny2019} and  \cite{sharma2021arXiv}, both of which uses random forests to classify high redshift merging galaxies. \cite{sny2019} classify mergers in a sample of $0.5 < z < 4.0$ Illustris-1 simulated HST galaxies, with a merger definition of $\pm 250$ Myr, and report a true positive rate of $\sim70\%$ across their redshift range. \cite{sharma2021arXiv} classify mergers in a sample of $0.5 < z < 3$ SPHGal simulated HST images, with a merger definition of $\pm 500$ Myr, and report a true positive rate of $0.95$ for the full data set (see their Figure 3).

There are a few possible explanations for the better performance of these studies. Both \cite{sny2019} and \cite{sharma2021arXiv} focused on merger classification performance, whereas here we try to optimize performance on both mergers and non-mergers. \cite{sharma2021arXiv} specifically note that their merger fraction is overestimated due to false positives. \cite{sny2019} trained random forests on single snapshots, whereas here we use redshift ranges, so the forests from \cite{sny2019} may be overtrained on a per-snapshot basis. Finally, \cite{sny2019} and \cite{sharma2021arXiv} use Illustris-1 and SPHGal, respectively, to create their simulated images rather than IllustrisTNG as we do here, and there may be differences between the three that make it easier or harder of select mergers with this technique.

\section{Summary and Conclusions} \label{sec:con}

In this work, we investigate using random forests to classify merging galaxies in simulated CEERS-like and nearly noiseless images, which were constructed from IllustrisTNG and the Santa Cruz SAM. We use the morphology programs \texttt{Galapagos-2} and \texttt{statmorph} to calculate a number of morphology parameters which were then used as inputs to the random forests. We also use IllustrisTNG merger history catalogs to define intrinsic merger labels which were also given to the forests. We train seven random forests for seven different redshift bins, and find the following results:
\begin{enumerate}
    \item The forests correctly classify $\sim$60\% of mock CEERS mergers and non-mergers across all redshift bins. The precision of the merger class increases with redshift while the precision of the non-merger class decreases with redshift. ROC curves and precision-recall curves indicate that the forests perform better than random classifiers. The random forests do not perform better when trained and tested on morphology parameters from nearly noiseless simulated images.
    \item Rest-frame asymmetry features tend to be most important for merger classifications at low redshift, while rest-frame bulge and clump features tend to be more important at higher redshifts.
    \item False positives tend to appear elongated with potential faint merger signatures, despite being no more likely to be closer to the $\pm250$ merger window than true negatives. False negatives tend to appear undisturbed by their recent mergers.
    \item Selecting different probability thresholds results in improved performance on the merger class at the cost of worse performance on the non-merger class.
    \item After correcting for the incompleteness and purity of the forests, we recover the true merger fraction of the mock CEERS dataset very well (using the $\pm 250$ Myr merger definition). The shape and slope of our mock CEERS corrected merger fraction and merger rate, $f_{\mathrm{merger}}$(RF) and $\mathbb{R}_{\mathrm{merger}}$(RF), match with theoretical Illustris predictions. However, our $f_{\mathrm{merger}}$(RF) and $\mathbb{R}_{\mathrm{merger}}$(RF) are underestimated compared to Illustris predictions.
\end{enumerate}

Given a sample of CEERS galaxies with unknown merger labels, the results of this work indicate that we could recover a reasonable merger fraction and merger rate. However, it would be difficult to disentangle specific true mergers from misclassified non-mergers.

One area of improvement lies with the segmentation map. The morphology parameters described in this work all depend on how galaxies are identified and deblended in the segmentation map, so great care must be taken in how sources are detected and deblended. The impact of source detection on merger identification is an important topic for future exploration.

Our findings suggest that we have reached the ceiling on how well random forests are able to identify mergers from these standard morphological parameters. Further improvement is likely to be gained through training convolutional neural networks (CNNs) to identify mergers directly from the images, which will be the subject of a future paper.

\begin{acknowledgments}

 Support for this work was provided by NASA through grants JWST-ERS-01345.015-A and HST-AR-15802.001-A awarded by the Space Telescope Science Institute, which is operated by the Association of Universities for Research in Astronomy, Inc., under NASA contract NAS 5-26555. This research is based in part on observations made with the NASA/ESA Hubble Space Telescope obtained from the Space Telescope Science Institute, which is operated by the Association of Universities for Research in Astronomy, Inc., under NASA contract NAS 5–26555. 

The authors acknowledge Research Computing at the Rochester Institute of Technology for providing computational resources and support that have contributed to the research results reported in this publication. \href{https://doi.org/10.34788/0S3G-QD15}{https://doi.org/10.34788/0S3G-QD15}

The authors acknowledge the Texas Advanced Computing Center (TACC) at The University of Texas at Austin for providing HPC resources that have contributed to the research results reported within this paper. \href{http://www.tacc.utexas.edu}{http://www.tacc.utexas.edu}

\end{acknowledgments}

\software{Source Extractor \citep{ber1996}, Galapagos-2 \citep{hau2013}, statmorph \citep{rod2019}}

\bibliography{rose_paper.bib}{}
\bibliographystyle{aasjournal}

\end{document}